\documentclass[pra,aps,showpacs,twocolumn,superscriptaddress,floatfix]{revtex4}

\usepackage{graphicx}
\usepackage{amsmath}
\usepackage{bm} 
\usepackage{color}
\usepackage{amssymb}

\newcommand{\beq}{\begin{eqnarray}}
\newcommand{\eeq}{\end{eqnarray}}
\newcommand{\bqa}{\begin{eqnarray}}
\newcommand{\eqa}{\end{eqnarray}}
\newcommand{\del}{\partial}

\newcommand{\gsim}{\hspace*{0.2em}\raisebox{0.5ex}{$>$}
     \hspace{-0.8em}\raisebox{-0.3em}{$\sim$}\hspace*{0.2em}}

\def\mqo2{{\!\!\!}}
\renewcommand{\vec}{\mathbf}

\begin{document}

%%%%%%%%%%%%%%%%%%%%%%%%%%%%%%%%%%%%%%%%
%The normal things to start with.
%%%%%%%%%%%%%%%%%%%%%%%%%%%%%%%%%%%%%%%%

\preprint{HISKP-TH-09/36}
\title{Three-body problem in heteronuclear mixtures with resonant interspecies interaction}
\author{K.\ Helfrich}%\email{helfrich@hiskp.uni-bonn.de}
\author{H.-W.\ Hammer}%\email{hammer@hiskp.uni-bonn.de}
\affiliation{Helmholtz-Institut f\"ur Strahlen- und Kernphysik (Theorie)\\
and Bethe Center for Theoretical Physics,
 Universit\"at Bonn, 53115 Bonn, Germany}
\author{D.~S.\ Petrov}
\affiliation{Laboratoire Physique Th\'eorique et Mod\`eles Statistique,
Universit\'e Paris Sud, CNRS, 91405 Orsay, France}
\affiliation{Russian Research Center Kurchatov Institute, Kurchatov
Square, 123182 Moscow, Russia\\}

\date{\today}

%%%%%%%%%%%%%%%%%%%%%%%%%%%%%%%%%%%%%%%%%
%The Abstract
%%%%%%%%%%%%%%%%%%%%%%%%%%%%%%%%%%%%%%%%%
\begin{abstract}
We use the zero-range approximation to study a system of two identical bosons interacting resonantly with a third particle. 
The method is derived from effective field theory. It reduces the three-body problem to an integral equation which we 
then solve numerically. We also develop an alternative approach which gives analytic solutions of the integral equation in 
coordinate representation in the limit of vanishing total energy. The atom-dimer scattering length, the rates of atom-dimer 
relaxation, and three-body recombination to shallow and to deep molecular states are calculated either analytically or 
numerically with a well-controlled accuracy for various energies as functions of the mass ratio, scattering length, and three-body parameter. 
We discuss in detail the 
relative positions of the recombination loss peaks, which in the universal limit depend only on the mass ratio. Our results 
have implications for ongoing and future experiments on Bose-Bose and Bose-Fermi atomic mixtures.
\end{abstract}
\maketitle

%%%%%%%%%%%%%%%%%%%%%%%%%%%%%%%%%%%%%%%%%
%Introduction
%%%%%%%%%%%%%%%%%%%%%%%%%%%%%%%%%%%%%%%%%
\section{Introduction}
\label{sec:intro}
The universal properties of particles with resonant
short-range interactions are a subject of intense research.
Such systems are characterized by a large scattering length
and display universal phenomena associated with a discrete scaling
symmetry \cite{Braaten:2004rn,Platter:2009gz}.
For identical bosons, Efimov found that there are
infinitely many trimer states with an accumulation point at the
scattering threshold when the $s$-wave scattering length $a$ is tuned
to the unitary limit $1/a =0$ \cite{Efimov70}:
%----------------------
\begin{eqnarray}
B^{(n)}_t = (e^{-2\pi/s_0})^{n-n_*} \hbar^2 \kappa^2_* /m,
\label{kappa-star}
\end{eqnarray}
%----------------------
where $m$ is the mass of the particles, $s_0 \approx 1.00624$, and
$\kappa_*$ is the binding wavenumber of the branch of Efimov states
labeled by $n_*$.  The geometric  spectrum in (\ref{kappa-star})
is the signature of a
discrete scaling symmetry with scaling factor $e^{\pi/s_0}\approx 22.7$.
For a finite scattering length larger than the range of the
interaction, the universal properties persist but
there is only a finite number of Efimov states.

Ultracold atoms are an ideal tool to study such phenomena since
the scattering length can be tuned experimentally using Feshbach
resonances. 
Efimov trimers in ultracold atomic gases can be observed
via their signature in three-body recombination rates
\cite{NM-99,EGB-99,BBH-00,BH01,Braaten:2003yc}. 
Kraemer {\it et al.}\ provided the first evidence for Efimov trimers 
in an ultracold gas of $^{133}$Cs atoms by observing the resonant 
enhancement of three-body recombination 
caused by the trimers \cite{Kraemer:2006}.
In a subsequent experiment with a mixture of $^{133}$Cs atoms and 
dimers, Knoop {\it et al.}\ observed a resonance in the loss 
of atoms and dimers \cite{Knoop:2008} which can be explained by an 
Efimov trimer crossing the atom-dimer threshold \cite{Helfrich:2009uy}.  

Several recent experiments have also obtained evidence of Efimov
physics with other bosonic atoms.

Zaccanti {\it et al.}\ measured the three-body recombination rate and
the atom-dimer loss rate in an ultracold gas of $^{39}$K atoms
\cite{Zaccanti:2008}.  They observed loss features at large positive and
negative values of the scattering length, positions of which agree with the discrete scaling symmetry.
Gross {\it et al.}\ measured the three-body recombination
rate in an ultracold system of $^7$Li atoms \cite{Gross:2009}.  They observed
a three-atom loss resonance and a three-body recombination minimum in the same 
universal region on different sides of a Feshbach resonance. Their 
positions are consistent with the universal predictions with
discrete scaling factor of 22.7. Using ultracold $^7$Li atoms as well,
Pollack {\it et al.}\ \cite{Hulet:2009} observed 11 three- and four-body 
loss features in the inelastic loss spectrum. Their relative locations
on either side agree well with the universal theory, while a systematic
deviation from universality appears when comparing features across 
the resonance. The origin of this deviation is not understood.
Barontini {\it et al.}\ \cite{Barontini:2009} investigated
the Bose-Bose mixture $^{41}$K-$^{87}$Rb and found three resonance
positions in the three-body loss. The two features for
negative scattering length were attributed to the two possible Efimov
trimers of the system, K-Rb-Rb and K-K-Rb,
hitting the three-atom threshold.

The Efimov effect can also occur for fermionic atoms with at least
three spin states. The first experimental
studies of many-body systems of $^6$Li atoms in the three lowest hyperfine
states were recently carried out by Ottenstein {\it et al.}\
\cite{Ottenstein:2008} and Huckans {\it et al.}\ \cite{Huckans:2008}. 
Theoretical
calculations of the three-body recombination rate supported the interpretation
that the narrow loss feature arises from an Efimov trimer crossing the 
three-atom
threshold \cite{Braaten:2008wd,NU:2009,Schmidt:2008fz,Wenz:2009}.  
Very recently,
another narrow loss feature was discovered in the region of much higher
magnetic fields by Williams {\it et al.}\ \cite{Williams:2009} and by Jochim and
co-workers \cite{Jochim}. In this region, the scattering length is
much larger and several recombination features have been 
predicted using the universal theory~\cite{Braaten:2009ey}.

In this paper, we focus on heteronuclear systems with two species of atoms where only the interspecies scattering length is large. 
For comparable masses the scaling factor is quite large (for equal masses $e^{\pi/s_0}\approx 1986.1$) as we now have 
only two resonant interactions out of three. However, in the case of
two heavy atoms and one light atom, this factor can become 
significantly smaller than the value 22.7 for identical bosons~\cite{Amado72,Efimov72,Efimov73,Braaten:2004rn}, which should stimulate experimental investigation of the 
discrete scaling invariance.  
Relaxation and recombination losses near an interspecies resonance have recently been investigated in mixtures of rubidium 
and potassium. The Bose-Fermi combination $^{87}$Rb-$^{40}$K has been
studied at the Joint Institute for Laboratory Astrophysics (JILA)~\cite{Zirbel} and measurements on the 
Bose-Bose mixture $^{41}$K-$^{87}$Rb have been carried out in Florence~\cite{Barontini:2009}. We apply our theory to 
these and other mixtures of interest for ongoing and planned experiments.

%%%%%%%%%%%%%%%%%%%%%%%%%%%%%%%%%%%%%%%%%
%Method
%%%%%%%%%%%%%%%%%%%%%%%%%%%%%%%%%%%%%%%%%
\section{Method}
\label{sec:method}
In this section, we set up the effective field theory method which provides a convenient implementation of the universal theory 
for large scattering length. We set $\hbar=1$ but restore the dimensions in our expressions for the recombination rate constants. 
We consider a system of one boson or fermion of mass $m_1$ (species 1) and two identical bosons of mass $m_2$ (species 2). We assume 
the interspecies interaction to be resonant and characterized by the $s$-wave scattering length $a\gg \ell_{vdW}$, where $\ell_{vdW}$ 
is the van der Waals range of the potential. Nonresonant intraspecies interaction will be neglected. If species 1 is also bosonic 
and weakly interacting, all the forthcoming results directly apply to the other possible (interacting) triple by simply exchanging 
the labels 1 and 2. We therefore include only the interaction between the atoms of species $2$ and the dimers. Hence, our effective 
Lagrangian reads
%-------------------
\beq
{\cal L}&=&\psi_1^{\dagger}\biggl(i\del_t+\frac{\nabla^2}{2 m_1}\biggr)\psi_1+\psi_2^{\dagger}
\biggl(i\del_t+\frac{\nabla^2}{2 m_2}\biggr)\psi_2 +g_2d^{\dagger}d
\nonumber\\
&&-g_2\Bigl(d^{\dagger}\psi_1\psi_2+\psi_1^{\dagger}\psi_2^{\dagger}d\Bigr)-\frac{g_3}{4}d^{\dagger}d\psi_2^{\dagger}\psi_2\, 
+\cdots,
\label{lagrangian}
\eeq
%---------------------
where the dots represent higher-order derivative interactions, and $g_2$ and $g_3$ are the bare two- and three-body coupling constants.

From the Lagrangian (\ref{lagrangian}), we can deduce Feynman rules and obtain the full dimer propagator and the three-body integral 
equation (see Ref.~\cite{Braaten:2004rn} for details on the derivation). For the full dimer propagator we get
%-------------------
\beq
\label{dimerpropagator}
D(P_0,\vec{P})=\frac{2\pi}{g_2^2\mu}\Biggl[\frac{1}{a}-\sqrt{-2\mu\Bigl(P_0-\frac{P^2}{2M}\Bigr)-i\epsilon}\Biggr]^{-1}\, ,
\eeq
%---------------------
where $P=|\vec{P}|$, $\mu=m_1m_2/(m_1+m_2)$ is the reduced mass, $M=m_1+m_2$ is the mass of the dimer, and the limit of $\epsilon\rightarrow +0$ is understood. The dimer wave function 
renormalization is given by $Z_D^{-1}=g_2^2a\mu^2/(2\pi)$. 

\begin{figure}[t]
   %     \vspace*{0.4cm}
	\centerline{\includegraphics*[width=0.8\hsize,clip]{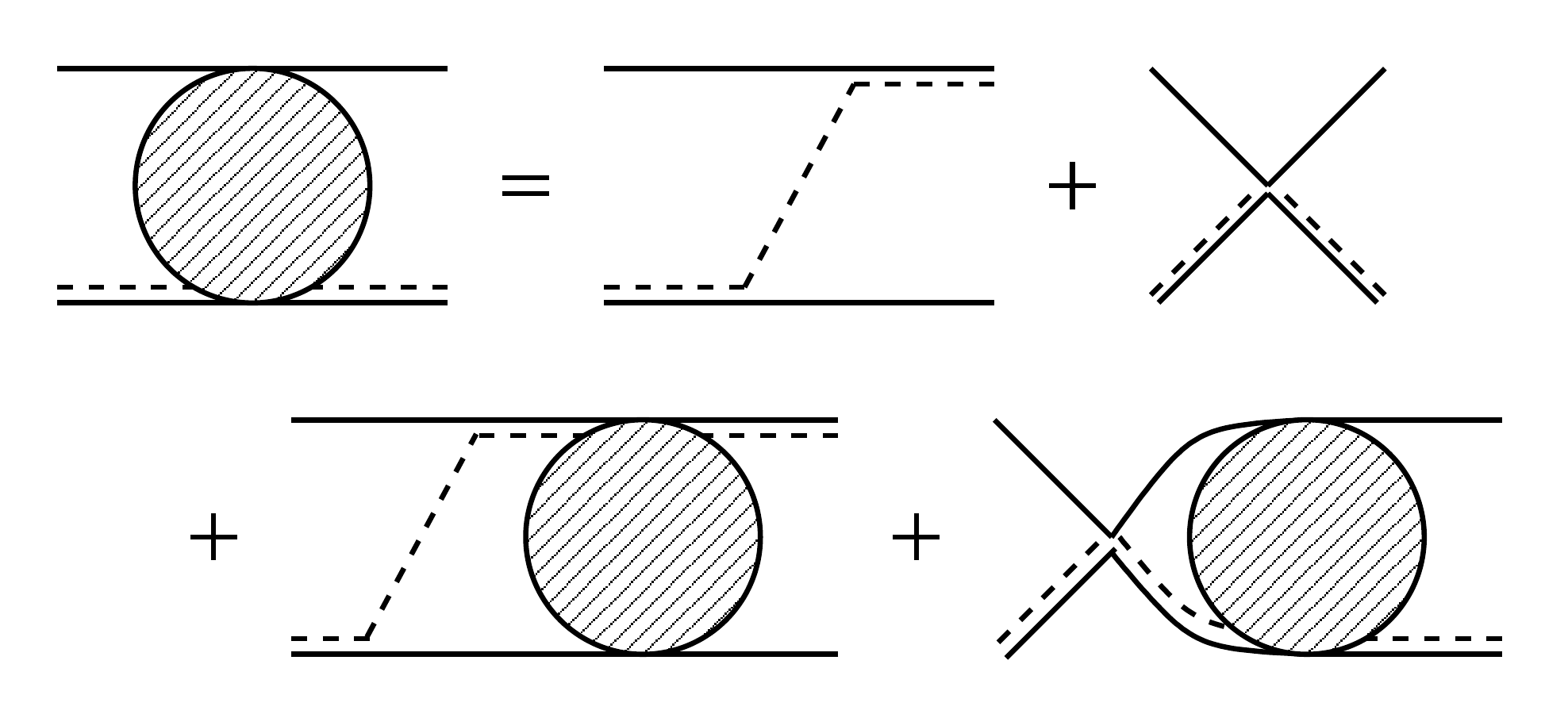}}
	\caption{Integral equation for the atom-dimer scattering amplitude ${\cal A}$. Solid (dashed) lines denote atom species $2$ ($1$). 
          Mixed double lines denote the full dimer propagator.}
	\label{fig:STM}
\end{figure}

The scattering between a dimer and an atom is described by the integral equation shown in Fig.~\ref{fig:STM}. Using the Feynman rules 
derived from Eq.~(\ref{lagrangian}) and projecting on relative $s$-waves, we have
%%%%%%%%%%%%%%%%%%%%%%%%%%%%%%%%%%%%%%%%%%%%%%%%%
%%%%%%%%%%%%%%%%%%%%%%%%%%%%%%%%%%%%%%%%%%%%%%%%%
%
%-------------------
\bqa
{\cal A}(p,k;E)&=&\frac{2\pi m_1}{a\mu^2}\Biggl[K(p,k)-\frac{g_3}{4m_1 g_2^2}\Biggr]\nonumber \\ 
&& +\frac{m_1}{\pi\mu}\int_0^{\Lambda} dq\, q^2\Biggl[K(p,q)-\frac{g_3}{4m_1 g_2^2}\Biggr]\nonumber \\
&&\;\times\frac{{\cal A}(q,k;E)}{-\frac{1}{a}+\sqrt{-2\mu\left(E-\frac{q^2}{2\mu_{AD}} \right)-i\epsilon}}\, ,\qquad
\label{STM}
\eqa
%---------------------
where $\mu_{AD}=m_2(m_1+m_2)/(2m_2+m_1)$ is the reduced mass of an atom and a dimer, the relative momenta of the incoming and 
outgoing atom-dimer pair are denoted by $p$ and $k$, respectively; and $E$ is the total energy. The contribution of the $s$-wave 
projected one-atom exchange is given by
%-------------------
\beq\label{Kernel}
K(p,q)=\frac{1}{2 p q}\ln\Biggl[\frac{p^2+q^2+2pq\frac{\mu}{m_1}-2\mu E-i\epsilon}{p^2+q^2-2pq\frac{\mu}{m_1}-2\mu E-i\epsilon}\Biggr]\, ,
\eeq
%---------------------
and the contribution of the three-body coupling $g_3$ can be written as
%-------------------
\beq
\frac{g_3}{4 m_1 g_2^2}=-\frac{H(\Lambda)}{\Lambda^2}\,,
\eeq
%---------------------
where $H(\Lambda)$ is a dimensionless log-periodic function of the cutoff $\Lambda$,
which depends on a three-body parameter $\Lambda_*$ \cite{Bedaque:1998kg}. 
The mass-ratio dependence of the discrete scaling factor $\exp(\pi/s_0)$ follows from the equation for $s_0$:
\beq
\label{s0equation}
s_0 \cosh(\pi s_0/2) -2\sinh(\phi s_0)/\sin(2\phi)=0\,,
\eeq
where we introduce the parameter
\beq\label{phi}
\phi=\arcsin\left[1/(1+\delta)\right]\,
\eeq
and the notation $\delta=m_1/m_2$. For particles of equal mass, the solution of Eq.~(\ref{s0equation}) is $s_0\approx 0.4137$ 
leading to the scaling factor $\exp(\pi/s_0)\approx 1986.1$.
Because of the log-periodicity of $H(\Lambda)$ one can always find a value of the cutoff $\Lambda$ with $H=0$. In practice, 
one can therefore simply omit the three-body coupling in the leading-order calculations and use the cutoff $\Lambda$ as a 
three-body parameter \cite{Hammer:2000nf}. We use this strategy in the following. For fixed $\delta$, the values of 
$\Lambda$ and $\Lambda_*$ are related by a multiplicative constant.

The scattering amplitude ${\cal A}$ has simple poles at the three-body bound-state energies $E=-B_t <0$. The energies can 
be obtained from the solution of the following homogeneous integral equation for the bound-state amplitude ${\cal B}$: 
%-------------------
\beq
\label{boundstateeq}
{\cal B}(p;B_t)&=&\frac{m_1}{\pi\mu}\int_0^{\Lambda}
\frac{dq\,q^2\,K(p,q)\;{\cal B}(q;B_t)}{-\frac{1}{a}+\sqrt{2\mu\left(B_t+\frac{q^2}{2\mu_{AD}}\right)}}\, ,
\eeq
%---------------------
which has nontrivial solutions only for three-body binding energies $B_t >0$.
In the following, we use Eqs.~(\ref{STM}) and(\ref{boundstateeq}) to describe three-body properties of heteronuclear mixtures.

%%%%%%%%%%%%%%%%%%%%%%%%%%%%%%%%%%%%%%%%%
%Results
%%%%%%%%%%%%%%%%%%%%%%%%%%%%%%%%%%%%%%%%%
\section{Numerical Results}
\label{sec:results}
Few-body loss phenomena offer a unique view on scattering processes in ultracold quantum gases. In particular, an enhancement of the 
loss rate can be an evidence of a few-body resonance. The universal theory predicts the relative positions of such resonances as a 
function of the scattering length. The universality can thus be tested experimentally by measuring the lifetime of a cold atomic gas 
as a function of $a$. Ideally, in order to see the universal scaling, one needs to detect more than one resonance in a single universal 
region, that is, a region where the three-body parameter can be assumed constant. This is believed to happen in a narrow vicinity of a 
Feshbach resonance, where large variations of $a$ are accompanied by (assumed) much weaker variations of the three-body parameter. 
We now discuss three-body loss resonances in a heteronuclear mixture as predicted by the universal theory.

%%%%%%%%%%%%%%%%%%%%%%%%%%%%%%%%%%%%%%%%%%%%%%%%%
\begin{figure}[t]
   %     \vspace*{0.4cm}
	\centerline{\includegraphics*[width=0.6\hsize]{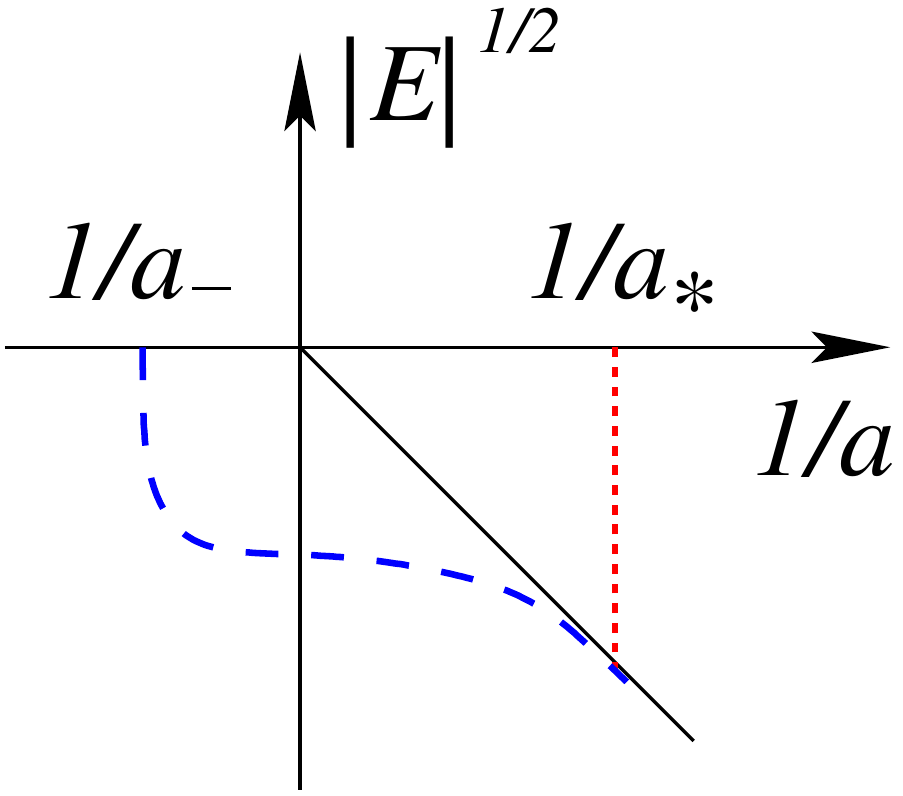}}
	\caption{(Color online) Dependence of the trimer energy on the inverse scattering 
                 length $1/a$ in arbitrary units (dashed line).
                 The parameters $a_-$ and $a_*$ specify where the trimer state hits the
                 three-atom and atom-dimer thresholds, respectively.}
	\label{fig:defasam}
\end{figure}
%%%%%%%%%%%%%%%%%%%%%%%%%%%%%%%%%%%%%%%%%%%%%%%%%
\subsection{Resonance positions}
The mechanism of three-body losses and its relation to the positions of Efimov levels in the heteronuclear case are qualitatively 
the same as for three identical bosons. The scattering-length dependence of the energy of a generic trimer is illustrated in 
Fig.~\ref{fig:defasam}. On the negative side of a Feshbach resonance the trimer hits the three-body scattering threshold at $a=a_-<0$, 
which leads to an enhanced probability of finding three atoms at distances of the order of $|a|$. Such atoms can then approach each 
other to distances of the order of $\ell_{vdW}$ and recombine into a deeply bound dimer and a residual atom. The released binding 
energy [of order $\hbar^2/(2\mu\ell_{vdW}^2)$] transforms into the kinetic energy of the recombination products, which hence leave 
the trap. On the positive side of the Feshbach resonance there exists a weakly bound (shallow) dimer state with binding energy 
$B_d=\hbar^2/(2\mu a^2)$. This formula taken with minus sign determines the atom-dimer threshold (solid line in Fig.~\ref{fig:defasam}). 
By following the dashed line in Fig.~\ref{fig:defasam} from negative
to positive values of $a$, one can see that the trimer crosses the atom-dimer threshold 
at $a=a_*>0$, where one predicts an elastic atom-dimer resonance. At this point formation of deep dimer states (in this case called relaxation) 
in atom-dimer collisions is also enhanced for the same reason as above. According to~\cite{Zaccanti:2008}, the atom-dimer 
scattering resonance should be noticeable even in a purely atomic sample due to rescattering processes. Indeed, before leaving the trap, 
shallow dimers formed in the process of three-body recombination can collide with other atoms. The recombination rate itself is 
featureless around $a=a_*$, but the atom-dimer cross section in the
vicinity of this point is highly $a$ dependent. Thus, at $a=a_*$ 
the three-body recombination can be enhanced in the sense that many more than three atoms are expelled from the trap leading 
to a measurable trap loss. We come back to this issue in Sec.~\ref{sec:comparison}. 

\begin{figure}[t]
   %     \vspace*{0.4cm}
	\centerline{\includegraphics*[width=0.95\hsize]{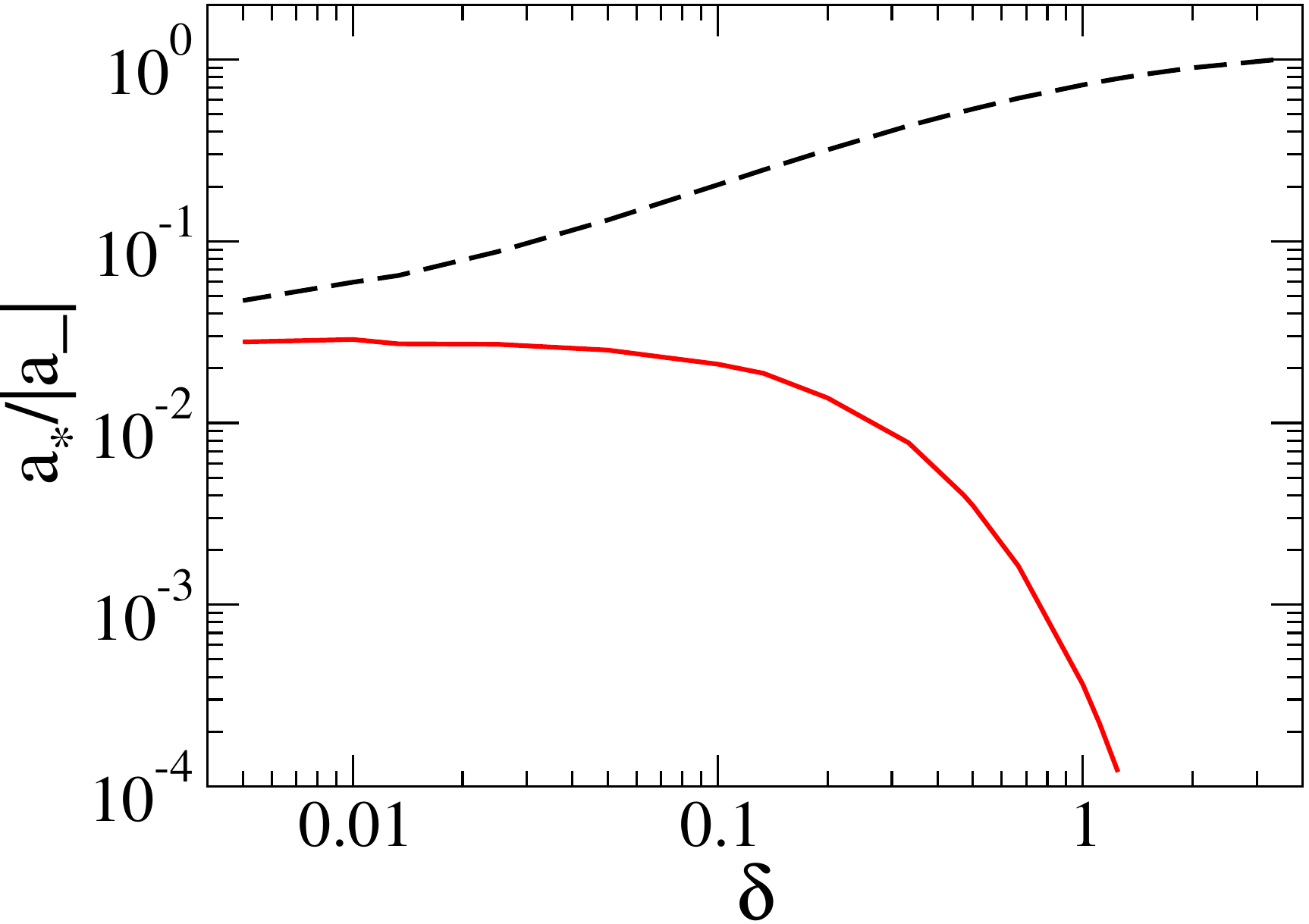}}
	\caption{(Color online) Solid line: $a_*^{(n)}/|a_-^{(n)}|$ vs
          $\delta$, where $n$ is the index of the Efimov state [see Eq.~(\ref{kappa-star})]. 
Dashed line: $a_*^{(n+1)}/|a_-^{(n)}|=\exp(\pi/s_0)a_*^{(n)}/|a_-^{(n)}|$.}
	\label{fig:ratios}
\end{figure}

The ratio of the two resonance positions, $a_*/|a_-|$, is of fundamental importance for studies of the universal three-body physics 
as in the universal limit it does not depend on the three-body parameter. In order to calculate this ratio we solve the bound-state 
Eq.~(\ref{boundstateeq}) for $B_t=0$,  $a<0$, and for $B_t=B_d$, $a>0$, with the same (arbitrary) cutoff $\Lambda$. The solid line 
in Fig.~\ref{fig:ratios} shows $a_*^{(n)}/|a_-^{(n)}|$ as a function of the mass ratio $\delta$. Here we use the index $n$ introduced 
in Eq.~(\ref{kappa-star}) in order to emphasize that the values of $a_*$ and $a_-$ are taken for one and the same Efimov state 
(connected by the dashed line in Fig.~\ref{fig:defasam}). The dashed line in Fig.~\ref{fig:ratios} differs from the solid one 
by the scaling factor $\exp(\pi/s_0)$ and shows the ratio $a_*^{(n+1)}/|a_-^{(n)}|$. Note that the scaling factor rapidly increases 
with $\delta$ for $\delta\gsim 1$ and one can conclude that a sequence of Efimov resonances is more likely to be seen in systems 
with smaller mass ratios.
%%%%%%%%%%%%%%%%%%%%%%%%%%%%%%%%%%%%%%%%%%%%%%%%%
%%%%%%%%%%%%%%%%%%%%%%%%%%%%%%%%%%%%%%%%%%%%%%%%%

\subsection{Three-body recombination for {\boldmath $a>0$}}
\begin{figure}[t]
   %     \vspace*{0.4cm}
  \centerline{\includegraphics*[width=0.8\hsize]{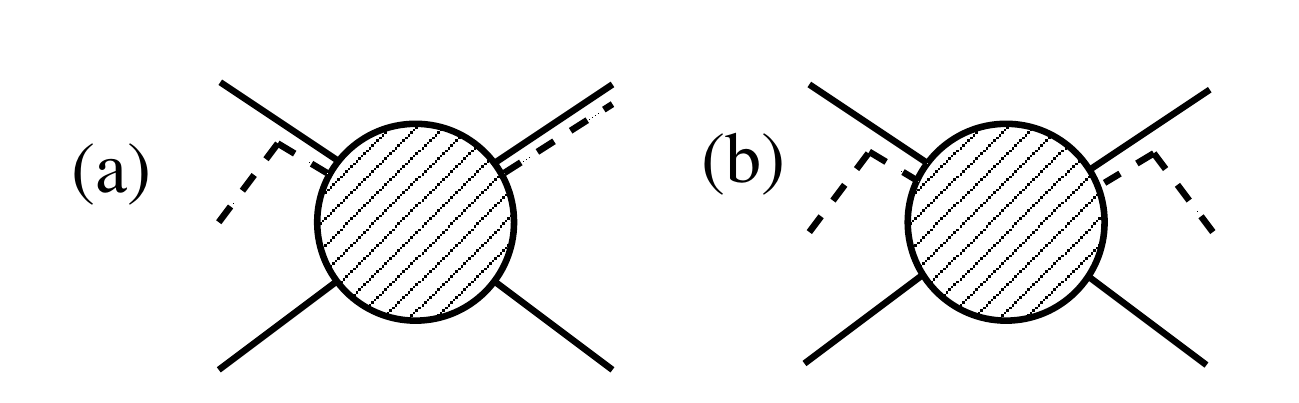}}
	\caption{Diagrammatic representation of (a) the three-body
          recombination amplitude and (b) the elastic three-body scattering. 
          Line patterns are the same as in Fig.~\ref{fig:STM}.}
	\label{fig:amplitudes}
\end{figure}

Let us now discuss the shapes of the inelastic loss resonances and calculate the three-body rate constants in a heteronuclear system. 
We first consider the case of positive scattering length, $a>0$, where the atoms can recombine into the shallow dimer and into deep 
dimers. The recombination into the shallow dimer can be related to the T-matrix element shown in Fig.~\ref{fig:amplitudes}(a). 
%%%%%%%%%%%%%%%%%%%%%%%%%%%%%%%%%%%%%%%%%%%%%%%%%
%%%%%%%%%%%%%%%%%%%%%%%%%%%%%%%%%%%%%%%%%%%%%%%%%
The event rate constant for inelastic scattering $\alpha$ is defined by the rate equation
%-------------------
\beq
\frac{d}{dt}n_2=2\frac{d}{dt}n_1=-2\alpha\,n_1n_2^2\, ,
\eeq
%---------------------
where $n_i$ denotes the atomic number densities of the corresponding species. 

\begin{figure}[t]
   %     \vspace*{0.4cm}
	\centerline{\includegraphics*[width=0.95\hsize]{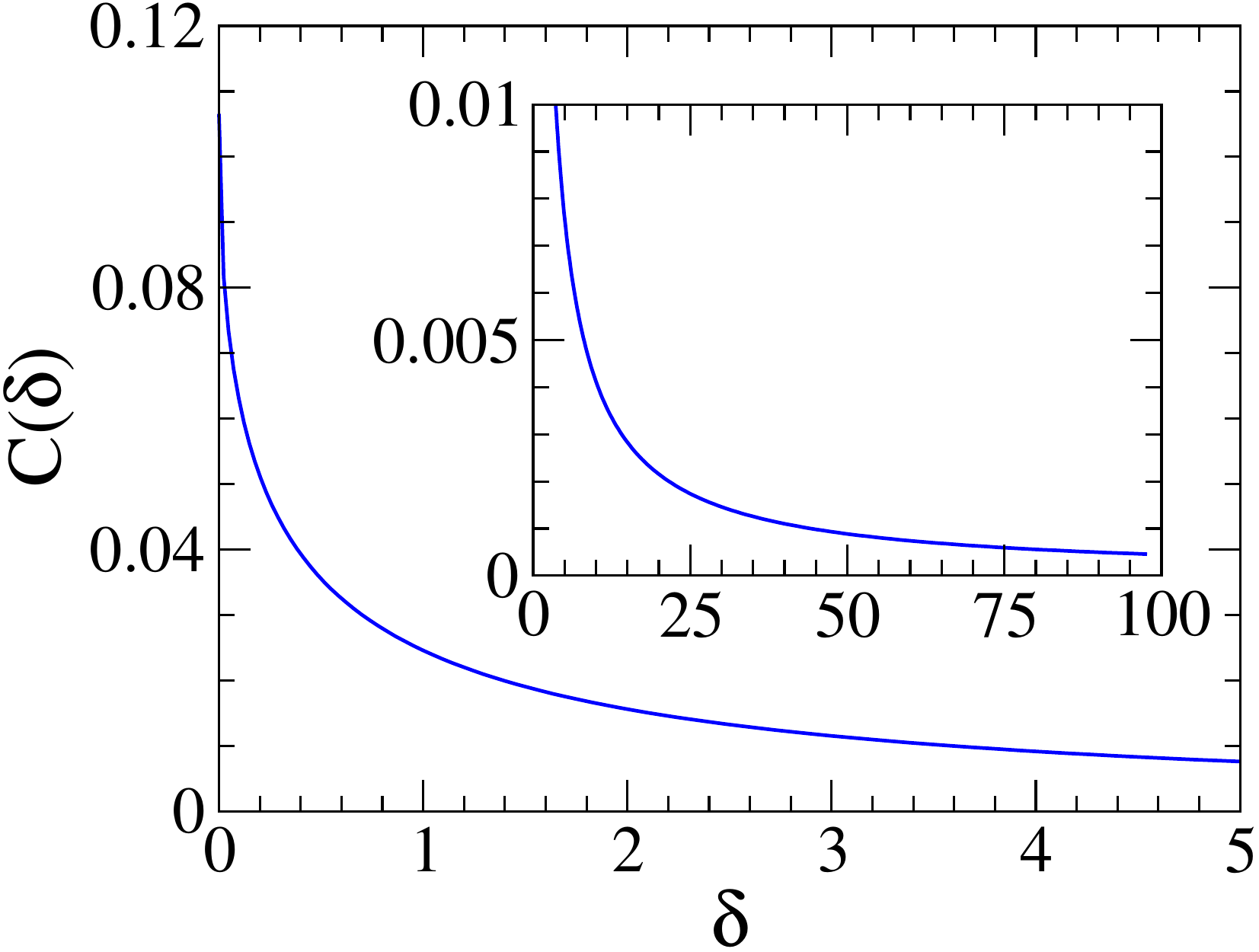}}
	\caption{(Color online) The coefficient $C$ for the different three-body recombination rates in dependence of the mass ratio 
$\delta=m_1/m_2$. The inset shows $C$ for larger values of $\delta$.
}
	\label{fig:coeff}
\end{figure}

The rate constant $\alpha_s$ for recombination into the shallow dimer is given by 
%-------------------
\beq
\label{eq:alphas}
\alpha_{s}=4\mu_{AD}\sqrt{\frac{\mu_{AD}}{\mu}}a^2\left|{\cal A}\left(0,k_D;0\right)\right|^2\, ,
\eeq
%---------------------
with the dimer breakup momentum $k_D=\sqrt{\mu_{AD}/\mu}\, a^{-1}$. 
If deep dimers are present, their effect on the recombination into the shallow
dimer can be incorporated by analytically continuing the three-body parameter into the 
complex plane~\cite{Braaten:2008wd}. We thus make the substitution
%-------------------
\beq
\Lambda\ \rightarrow\ \Lambda\exp(i\eta_*/s_0)
\label{eq:reprule}
\eeq
%---------------------
in Eq.~(\ref{STM}), where $\eta_*$ accounts for the effect of the deep dimers. 
A nonzero value of $\eta_*$ also generates the width of the Efimov trimers. By evaluating Eq.~(\ref{eq:alphas}) numerically, 
we find that the known analytical formula 
for the three-boson case~\cite{Braaten:2009ey} simply acquires a new mass-dependent overall coefficient. The modified analytical 
formula is hence
%-------------------
\beq
\alpha_{s}=C(\delta)\,\frac{D(\sin^2[s_0\ln(a/a_{0*})]+\sinh^2\eta_*)}{\sinh^2(\pi s_0+\eta_*)+\cos^2[s_0\ln(a/a_{0*})]}\frac{\hbar a^4}{m_1},
\label{ashallow}
\eeq
%---------------------
where $D=128\pi^2(4\pi-3\sqrt{3})$ and the mass-dependent coefficient is denoted by $C(\delta)$.  The parameter $a_{0*}$ 
gives the position of the minimum in the three-body recombination.
The coefficient $C(\delta)$ is shown in Fig.~\ref{fig:coeff}.  The error in the extraction of $C(\delta)$ from fitting 
Eq.~(\ref{ashallow}) to our numerical results for $\alpha_s$ is of order $10^{-3}$ for  $\delta \leq 2$. For larger values 
of $\delta$ the numerical extraction of $C$ becomes 
difficult because of a very large value of the scaling factor. To depict $C(\delta)$ for $\delta \geq 2$ we use 
the analytical formula (\ref{Cdelta}) derived in Sec.~\ref{sec:anaresults}. 
%%%%%%%%%%%%%%%%%%%%%%%%%%%%%%%%%%%%%%%%%%%%%%%%%
%%%%%%%%%%%%%%%%%%%%%%%%%%%%%%%%%%%%%%%%%%%%%%%%%

Although our calculations in this section are conducted by varying the complex three-body parameter $\Lambda$, 
we present the results in terms of the practically relevant length
parameters $a_-<0$, $a_*>0$, and $a_{0*}>0$, and the 
dimensionless elasticity parameter $\eta_*$. The universal theory predicts that the ratios $a_*/|a_-|$ and $|a_-|/a_{0*}$ 
depend only on the mass ratio $\delta$. The former is shown in Fig.~\ref{fig:ratios} and the latter is 
$|a_-|/a_{0*}=\exp(\pi/2s_0)$ as is derived in Sec.~\ref{sec:anaresults}. This fixes the relative positions of all the 
three-body loss features on both sides of the Feshbach resonance. %(if plotted versus $\log |a|$). 

The total rate of three-body recombination into all dimers (shallow and deep) for $a>0$ can be obtained from the optical 
theorem. It relates the imaginary part of the forward T-matrix 
element [shown in Fig.~\ref{fig:amplitudes}(b)] for vanishing momenta to the event rate constant of inelastic scattering, $\alpha$. 
This leads to the total recombination rate constant
%-------------------
\beq\label{alpha_tot}
\alpha_s+\alpha_d={\rm Im} T_{122\rightarrow 122}=8\pi a^3 {\rm Im}\bar{{\cal A}}(0,0;0)\, ,
\eeq
%---------------------
where $\bar{{\cal A}}$ denotes the appropriately infrared subtracted amplitude \cite{Braaten:2009ey}:
%-------------------
%\begin{widetext}
%\begin{equation}
%\bar{{\cal A}}(p,k;E)
%={\cal A}(p,k;E)-\frac{4\pi(1+\delta)}{m_1 a p^2}+\frac{4\pi (1+\delta)^2}{m_1 p}\arcsin\frac{1}{1+\delta}+
%\frac{8a}{m_1}\left[(1+\delta)^2\arcsin\frac{1}{1+\delta}-\sqrt{\delta(2+\delta)}\right]\ln p\, .
%\label{eq:sub}
%\end{equation}
%\end{widetext}
%-------------------
\bqa
&&\bar{{\cal A}}(p,k;E)
={\cal A}(p,k;E)-\frac{4\pi(1+\delta)}{m_1 a p^2}\nonumber\\
&&\quad+\frac{4\pi (1+\delta)^2}{m_1 p}\arcsin\left[1/(1+\delta)\right]
\nonumber\\
&&\quad+\frac{8a}{m_1}\left[(1+\delta)^2\arcsin\left[1/(1+\delta)\right]-\sqrt{\delta(2+\delta)}\right]\ln p\, .
\nonumber\\
\label{eq:sub}
\eqa
By subtracting Eq.~(\ref{ashallow}) from Eq.~(\ref{alpha_tot}) we find the rate constant for the recombination into deep dimers:
%-------------------
\beq
\label{adeep2}
\alpha_d=C(\delta)\frac{D\coth(\pi s_0)\cosh(\eta_*)\sinh(\eta_*)}{\sinh^2(\pi s_0+\eta_*)+\cos^2[s_0\ln(a/a_{0*})]}
\frac{\hbar a^4}{m_1}\, ,
\eeq
%---------------------
where the coefficients $C(\delta)$ and $D$ are the same as in Eq.~(\ref{ashallow}). 

When $s_0$ is not too small such that $\exp(2\pi s_0)\gg 1$ (see Table~\ref{tab:nr1}), the denominators in 
Eqs.~(\ref{ashallow}) and (\ref{adeep2}) are practically independent
of $a$. In this case the $a$ dependence of $\alpha_s$ 
and $\alpha_d$ is simplified, and the corresponding expressions are known in the case of three identical bosons 
(see, for example, Ref.~\cite{Braaten:2004rn}).                   

\subsection{Atom-dimer scattering}
\begin{figure}[t]
    \centerline{\includegraphics*[width=0.95\hsize]{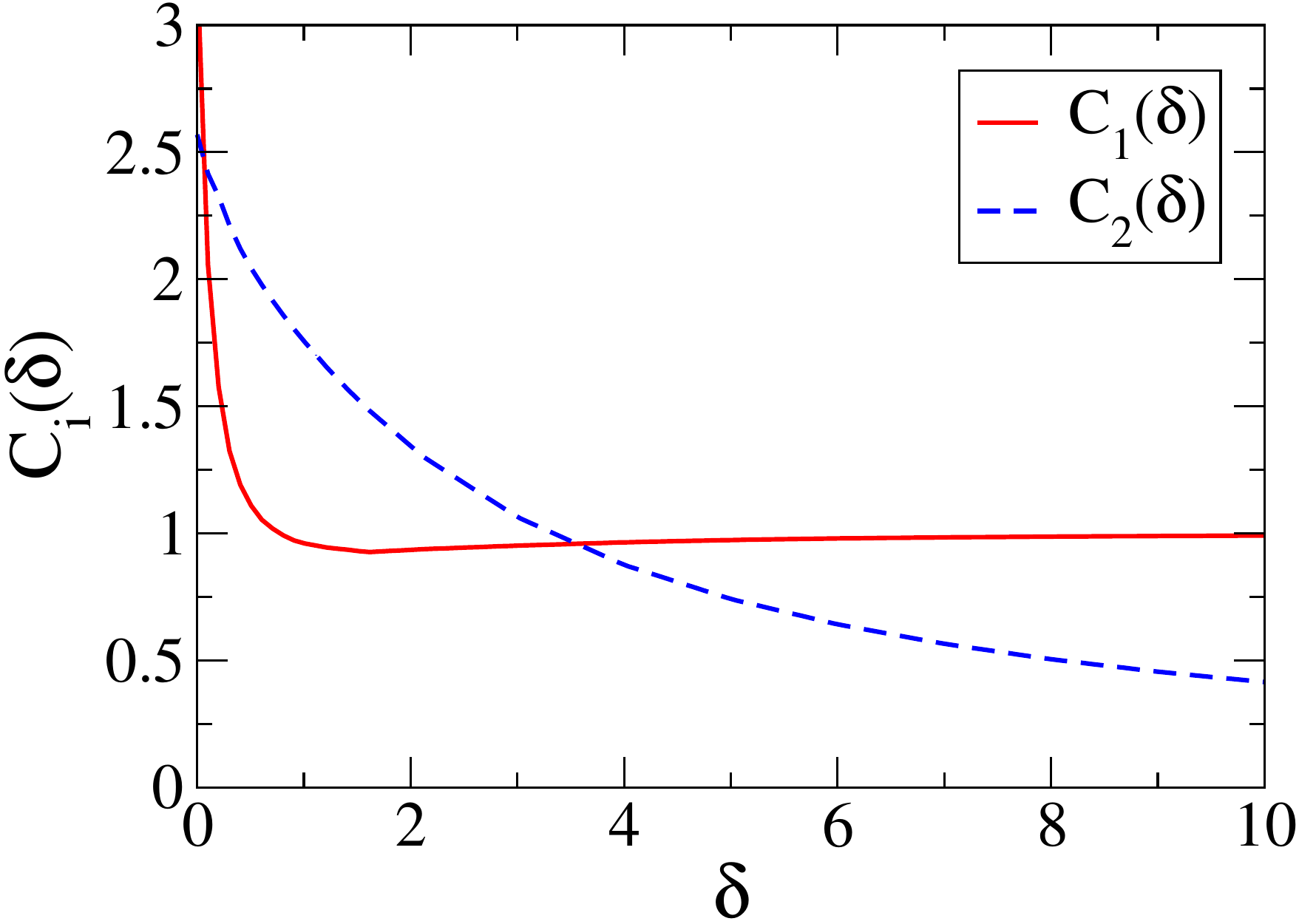}}
	\caption{(Color online) The parameters $C_1(\delta)$ and
          $C_2(\delta)$ in the expression for the atom-dimer
          scattering length, Eq.~(\ref{aADeq}).}
	\label{fig:coeffaAD}
\end{figure}

On the positive side of the Feshbach resonance ($a>0$) it is also possible to prepare an ultracold mixture of atoms and 
weakly bound dimers (see, for example, Refs.~\cite{Zirbel,Knoop:2008}). An important observable in this case is the 
atom-dimer scattering length. Within our theory, it is given by  
%-------------------
\beq
a_{AD}=-\frac{\mu_{AD}}{2\pi}{\cal A}\left(0,0;-\frac{1}{2\mu a^2}\right)\, ,
\eeq
%---------------------
and its universal dependence on $a$ is parametrized by
%-------------------
\beq
\label{aADeq}
a_{AD}=\Bigl(C_1(\delta)+C_2(\delta)\cot[s_0\ln(a/a_*)]\Bigr)a\, ,
\eeq
%---------------------
where the coefficients $C_1(\delta)$ and $C_2(\delta)$, calculated numerically, are shown in Fig.~\ref{fig:coeffaAD}. 
Here we estimate the numerical error in the determination of $C_1(\delta)$ and $C_2(\delta)$ to be of order $10^{-3}$.
%%%%%%%%%%%%%%%%%%%%%%%%%%%%%%%%%%%%%%%%%%%%%%%%%
%%%%%%%%%%%%%%%%%%%%%%%%%%%%%%%%%%%%%%%%%%%%%%%%%

Efremov and collaborators have recently derived Eq.~(\ref{aADeq}) for the atom-dimer scattering length in the Born-Oppenheimer 
approximation valid in the limit $\delta\rightarrow 0$ \cite{Efremov:09}. For $\delta=0.081$, corresponding to the 
$^7$Li-$^{87}$Rb-$^{87}$Rb system, our values for the coefficients
$C_1$ and $C_2$ agree with the ones given in Ref.~\cite{Efremov:09}  
to within 2-3$\%$ (see Table~\ref{tab:nr1}). However, we observe a stronger discrepancy in between our value, $s_0 = 1.523$, 
and the Born-Oppenheimer result, $s_0 = 1.322$, for this system \cite{Efremov:09}.
%(note the reversed mass ratio), our values: $C_1=2.22$, $C_2=2.47$ and we used $s_0=1.52$($\exp(\pi/s_0)=7.9$), 
%Efremov's value: $C_1=2.17$, $C_2=2.55$ and $s_0=1.322$,$\exp(\pi/s_0)=10.8$. 

The effect of deep dimers on the atom-dimer scattering process can be incorporated by replacing 
$a_*\, \rightarrow\, a_* \exp(-i\eta_*/s_0)$, equivalent to Eq.~(\ref{eq:reprule}). 
At the scattering threshold,
the atom-dimer relaxation rate constant $\beta$, defined by the rate equation
%---------------------
\beq
\frac{d}{dt}n_A=\frac{d}{dt}n_D=-\beta n_A n_D\,,
\eeq
%--------------------- 
is given by \cite{Braaten:2003yc}
%---------------------
\begin{eqnarray}
&&\hspace{-0.8cm}\beta(E=-B_d)=-(4\pi\hbar/\mu_{AD})\;{\rm Im} a_{AD}\nonumber\\
&&\hspace{-0.5cm}
= 2\pi C_2(\delta)\frac{\delta(\delta+2)}{\delta+1}\frac{\sinh(2\eta_*)}{\sin^2[s_0\ln(a/a_*)]+\sinh^2 \eta_*}
\frac{\hbar a}{m_{1}}\, .\nonumber\\
\label{eq:beta}
\end{eqnarray}
%---------------------

%
Furthermore, we can calculate the atom-dimer relaxation rate constant 
above threshold. It is related to the inelastic atom-dimer scattering cross section by
%-------------------
\beq
\beta(E)=\frac{k}{\mu_{AD}}\sigma^{({\rm inel.})}_{AD}(E)\, ,
\eeq
%-------------------
where $k=\sqrt{2\mu_{AD}(E+B_d)}$. The energy dependent inelastic cross section is given by the difference of the total 
and the elastic cross sections,
%-------------------
\beq
\sigma^{({\rm inel.})}_{AD}(E)=\frac{2\mu_{AD}}{k}{\rm Im}{\cal A}(k,k;E)-\frac{\mu_{AD}^2}{\pi}\left|{\cal A}(k,k;E)\right|^2\,.
\nonumber\\
\eeq
%-------------------
We can use this formula up to the dimer breakup threshold at $k=k_D$
and thus map out the trajectory of the resonance peak. 
It moves from $a_*$ at the scattering threshold, $E=-B_d$, to $|a_-|$ at 
the dimer breakup threshold, $E=0$. For $\delta<3.475$ the resonance peak moves to values $a>a_*$.
Starting at $\delta=3.475$, where we have exactly $a_*/|a_-|=1$, it reverses this behavior and moves to  values $a<a_*$.
The peak height diminishes considerably with the energy. This effect is 
very large, especially for small values of $\delta$. For example, for $\eta_*=0.1$ and $\delta=0.1$, the peak at $E=0$ is smaller 
by a factor of 706 than the peak at $E=-B_d$. For $\delta=10$ this ratio still is 19.1. At $E=0$, we find excellent agreement with 
the analytical formula
%------------------
\begin{eqnarray}
\label{eq:beta_e0}
&&\hspace{-1cm}\beta(E=-0)=\pi [\delta(\delta+2)]^{3/2}/(\delta+1)^2\nonumber\\
&&\hspace{-0.5cm}\times \frac{\sinh(2\pi s_0)\sinh(2\eta_*)}
{\sinh^2(\pi s_0+\eta_*)+\cos^2[s_0\ln(a/a_{0*})]}\frac{\hbar a}{m_1}\,,
\end{eqnarray}
%-------------------
which is derived in Sec.~\ref{sec:anaresults}.
The peak position of $\beta\, (E=-0)$ coincides exactly with the position of the maximum of the three-body recombination rate at threshold.

\begin{figure}[t]
    \centerline{\includegraphics*[width=0.95\hsize,clip]{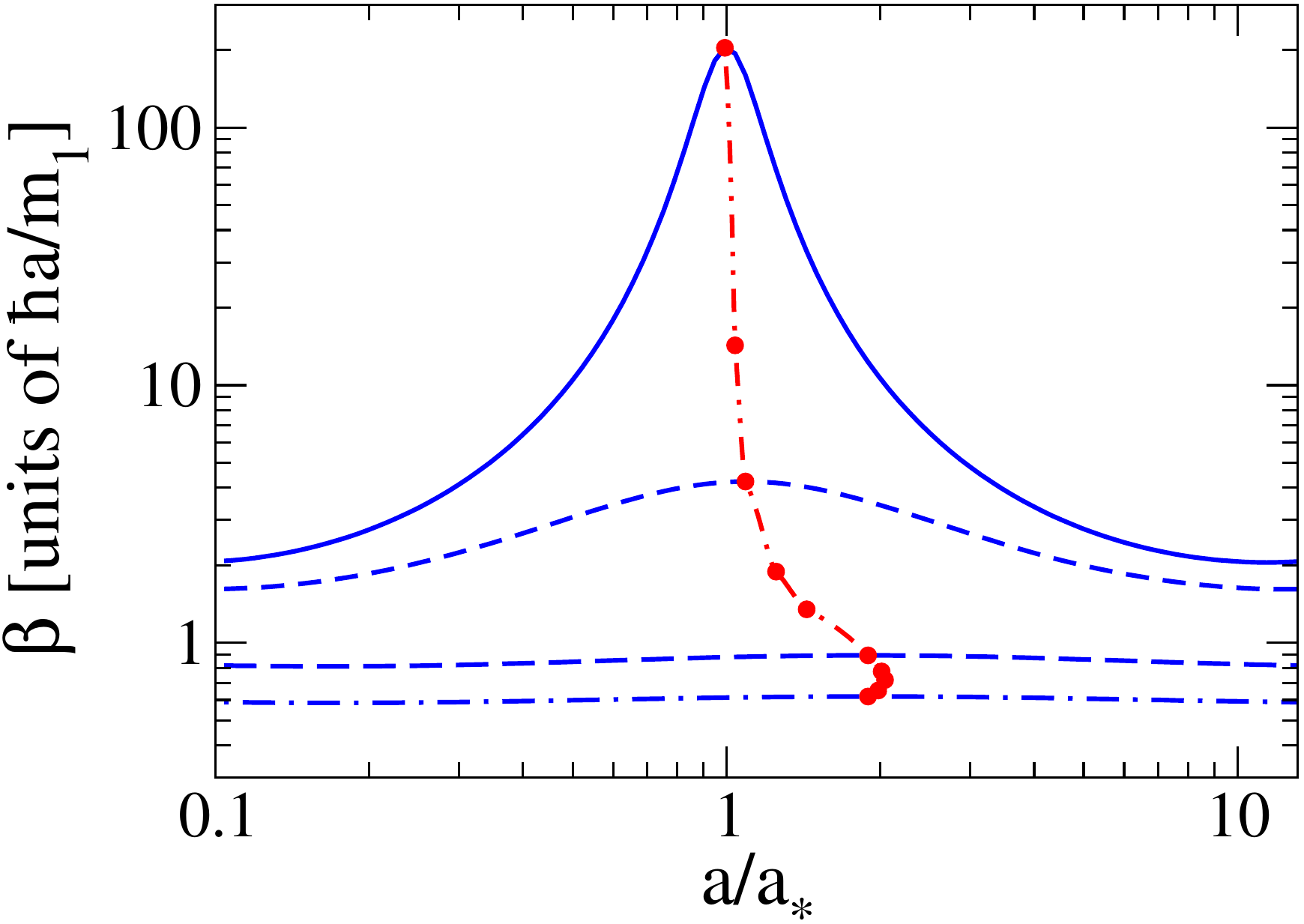}}
	\caption{(Color online) The dimer relaxation rate constant $\beta$ in units of $\hbar a/m_1$ for $\eta_*=0.1$
and $\delta=0.471$ as function of $a/a_*$. The solid, short-dashed, long-dashed, and dot-dashed lines
show $\beta$ for $E/B_d= -1,\, -0.95,\, -0.5$, and 0, respectively. The double-dot-dashed
line indicates the trajectory of the resonance maximum as the energy is increased from 
$-B_d$ to zero.}
	\label{fig:beta2}
\end{figure}

In Fig.~\ref{fig:beta2}, we show numerical results for 
$\beta$ for $\eta_*=0.1$ and $\delta=0.471$ corresponding to the K-Rb-Rb 
system observed in the Florence experiment~\cite{Barontini:2009}. 
%%%%%%%%%%%%%%%%%%%%%%%%%%%%%%%%%%%%%%%%%%%%%%%%%
%%%%%%%%%%%%%%%%%%%%%%%%%%%%%%%%%%%%%%%%%%%%%%%%%
The solid, short-dashed, long-dashed, and dot-dashed lines
show $\beta$ for $E/B_d= -1,\, -0.95,\, -0.5$, and 0, respectively. As the energy is increased toward
the breakup threshold, the resonance height decreases strongly and the resonance becomes less
pronounced. The double-dot-dashed
line shows the trajectory of the resonance maximum as the energy is increased from 
$-B_d$ to zero. As the energy is increased, the resonance position is shifted from $a_*$ toward
larger values of $a$ until it reaches its maximum value of $2.04 a_*$ for $E/B_d\approx-0.25$. For larger
energies, the resonance position moves back to smaller values of $a$ and reaches $|a_-|=1.89 a_*$ at
the dimer breakup threshold.  

\subsection{Three-body recombination for {\boldmath $a<0$}}
On the negative side of the Feshbach resonance, shallow dimers are absent and atoms can only recombine into deep dimers. 
The corresponding rate constant is again determined by using the optical theorem
%-------------------
\beq
\label{eq:alphad}
\alpha_{d}={\rm Im} T_{122\rightarrow 122}=8\pi a^3 {\rm Im}\bar{{\cal A}}(0,0;0)\,.
\eeq
%-------------------
We have performed numerical calculations of $\alpha_{d}$ for mass ratios $\delta \leq 2$, where 
the numerical accuracy is better than $0.1\%$. Our results agree with the formula 
%-------------------
\beq
\label{adeep}
\alpha_d=\frac{C(\delta)}{2}\frac{D\coth(\pi s_0) \sinh(2\eta_*)}{\sin^2[s_0\ln(a/a_-)]+\sinh^2(\eta_*)}\frac{\hbar a^4}{m_1}\, ,
\eeq
%---------------------
where the coefficients $C(\delta)$ and $D$ are the same as in Eqs.~(\ref{ashallow}) and (\ref{adeep2}). 
Equation~(\ref{adeep}) is derived in the next section.

%%%%%%%%%%%%%%%%%%%%%%%%%%%%%%%%%%%%%%%%%
%Results
%%%%%%%%%%%%%%%%%%%%%%%%%%%%%%%%%%%%%%%%%
\section{Analytical Approach}
\label{sec:anaresults}

Relying on an analysis of the recombination process in configuration space, Macek {\it et al.} \cite{Macek:2005,Macek:2006}, 
Petrov \cite{Petrov-octs}, and Gogolin with co-workers \cite{Gogolin:2008} have obtained a completely analytic expression for the 
three-body recombination rate constant for identical bosons with a
large positive scattering length $a$ when there are no deep dimers (i.e., $\eta_*=0$). Braaten and Hammer generalized this result to finite $\eta_*$ by making the analytic continuation 
$\kappa_* \to \kappa_* \exp (i \eta_* / s_0)$ in the amplitude of this process \cite{Braaten:2004rn}.  

In this section we would like to present an approach that allows one to calculate almost all zero-energy three-body observables 
analytically [in particular, $\alpha_s$, $\alpha_d(a>0)$, and $\alpha_d(a<0)$] for heteronuclear Bose-Fermi or Bose-Bose mixtures. 
We generalize the method introduced in Ref.~\cite{Petrov3Fermions} for fermionic mixtures and present it in more detail, 
also because it is conceptually different from the approaches of Refs.~\cite{Macek:2005,Macek:2006,Gogolin:2008}.

Let us consider the process of three-body recombination into the shallow heteronuclear dimer. When the total energy of the system, 
$E$, becomes smaller than the binding energy of the dimer, the three-body recombination rate tends to a finite zero-energy value. 
In order to calculate this quantity we work in the coordinate representation. It is instructive to derive the real-space 
analog of Eq.~(\ref{STM}) directly from the Schr\"odinger equation. At $E=0$ the latter reads
\begin{equation}\label{Schr}
\left[ \sum_{i=1,2}\left(-\frac{\hbar^2\nabla_{\bf R_i}^2}{2m_2}+U(|{\bf r -R_i}|)\right)-\frac{\hbar^2\nabla_{\bf r}^2}{2m_1}\right] \Psi=0,
\end{equation}  
where ${\bf R_i}$ are the coordinates of the bosons and ${\bf r}$ is the coordinate of the third atom. For the interspecies interaction 
we use the zero-range Fermi pseudopotential
\begin{equation}\label{pseudopot}
U(|{\bf r -R_i}|) = \frac{2\pi\hbar^2 a}{\mu} \delta ({\bf r -R_i})\frac{\partial}{\partial |{\bf r -R_i}|}( |{\bf r -R_i}| \cdot).
\end{equation}

Let us temporarily adopt the units $\hbar=2\mu=a=1$, separate the center-of-mass motion, and introduce the new coordinates 
${\bf x}=[{\bf R_1}\sin\phi-{\bf R_2}+{\bf r}(1-\sin\phi)]/\cos\phi$ and ${\bf y}={\bf r-R_1}$, where $\phi$ is defined 
in Eq.~(\ref{phi}). The bosonic symmetry condition, $\Psi({\bf R_1,R_2,r})=\Psi({\bf R_2,R_1,r})$, in the new coordinates reads
\begin{equation}\label{symmetry}
\Psi({\bf x,y})=\Psi(-{\bf x}\sin\phi+{\bf y}\cos\phi,{\bf x}\cos\phi+{\bf y}\sin\phi),
\end{equation}
and Eq.~(\ref{Schr}) takes the form
\begin{eqnarray}\label{SchrNew}
&&\hspace{-1.1cm}(-\nabla_{\bf x}^2-\nabla_{\bf y}^2)\Psi({\bf x,y})=f({\bf x})\delta ({\bf y})\nonumber\\
&&+f(-{\bf x}\sin\phi+{\bf y}\cos\phi)\delta ({\bf x}\cos\phi+{\bf y}\sin\phi),
\end{eqnarray}  
where we define
\begin{equation}\label{f}
f({\bf x}):=-4\pi \lim_{y\rightarrow 0} \partial[y\Psi({\bf x,y})]/\partial y.
\end{equation}
From Eqs.~(\ref{SchrNew}) and (\ref{f}) it is easy to see that
\begin{equation}\label{BethePeierls}
\Psi({\bf x,y})\xrightarrow[y\rightarrow 0]{}f({\bf x})(1/y-1)/4\pi.
\end{equation}
Thus, the function $f$ describes the motion of an atom relative to the other two atoms when they are on top of each other. 
In some cases it is useful to consider $f$ as an atom-dimer wave function. Indeed, for large $x$ expressions 
$\Psi\propto \phi_b(y)\exp(ix)/x$ and $f(x)\propto (8\pi)^{1/2}\exp(ix)/x$ equivalently describe the relative outgoing 
motion of the products of three-body recombination. The coefficient $(8\pi)^{1/2}$ is obtained by comparing 
Eq.~(\ref{BethePeierls}) and the small-$y$ asymptote of the normalized
molecular wave function
\begin{equation}\label{phi_b}
\phi_b(y)=\exp(-y)/\sqrt{2\pi}y\xrightarrow[y\rightarrow 0]{}(1/y-1)/\sqrt{2\pi}.
\end{equation}
Therefore, in order to calculate the three-body recombination rate
constant we first derive and solve an equation for $f$, 
then separate the large-$x$ asymptote $f\propto \exp(ix)/x$, and finally relate the coefficient in front of it to the 
three-body recombination rate constant. 

Using the Green's function $G(X)=1/4\pi^3X^4$ of Eq.~(\ref{SchrNew}) we can express $\Psi$ through $f$:
\begin{eqnarray}\label{PsiThroughf}
&&\hspace{-1.1cm}\Psi({\bf x,y})=\Psi_0 + \int [G(\sqrt{({\bf x-x'})^2+y^2})\nonumber\\
&&\hspace{-0.8cm}+G(\sqrt{({\bf x+x'}\sin\phi)^2+({\bf y-x'}\cos\phi)^2})]f({\bf x'})d^3 x',
\end{eqnarray} 
where $\Psi_0$ is a solution of the homogeneous Euler equation $(-\nabla_{\bf x}^2-\nabla_{\bf y}^2)\Psi_0=0$ without singularities. 
In the case of three-body recombination, $\Psi_0$ is a correctly
normalized wave function of three free atoms. We consider the atoms in a volume $V$ in a state where the two bosons are not in the same quantum state 
(cold thermal gas). Then, in the region relevant for the recombination ($x\lesssim 1$, $y \lesssim 1$), we get
\begin{equation}\label{Psi_0}
\Psi_0=\sqrt{2}V^{-3/2}.
\end{equation}
We point out that $\Psi_0$ should be set to zero for the problem of atom-dimer scattering just below the breakup 
threshold ($E=-0$) as in this case the atoms cannot move freely at large distances.

The equation for $f$ is obtained by substituting Eq.~(\ref{PsiThroughf}) into Eq.~(\ref{f}). We write it in the form
\begin{equation}\label{CoorSTM}
(\hat L -1)f({\bf x})=4\pi \Psi_0,
\end{equation} 
where the integral operator $\hat L$ is defined as
\begin{eqnarray}\label{L}
&&\hspace{-1.1cm}\hat L f({\bf x})=4\pi \int \{ G(|{\bf x-x'}|)[f({\bf x})-f({\bf x'})]\nonumber\\
&&\hspace{-0.8cm}-G(\sqrt{x^2+x'^2+2{\bf xx'}\sin\phi})f({\bf x'})\}d^3 x'.
\end{eqnarray} 
The operator $\hat L$ conserves angular momentum, and Eq.~(\ref{CoorSTM}) can be written as a set of uncoupled equations 
for each spherical harmonic of $f({\bf x})$. For the three-body
recombination problem at hand we look for a spherically 
symmetric solution $f({\bf x})=f(x)$. Then, Eq.~(\ref{L}) reduces to
\begin{eqnarray}\label{L0}
\hat L f(x)&=&\frac{4}{\pi} \int_0^\infty \left[\frac{f(x')-f(x)}{(x^2-x'^2)^2}\right.\nonumber\\
&+&\left.\frac{f(x')}{(x^2+x'^2)^2-4x^2x'^2\sin^2 \phi}\right] x'^2 d x'.
\end{eqnarray} 
The integrals in Eqs.~(\ref{L}) and (\ref{L0}) are taken as principal values \cite{06PRA}.

Let us now discuss the structure of possible solutions of Eq.~(\ref{CoorSTM}). Obviously, $f(x)$ is a sum of a particular 
solution of the inhomogeneous Eq.~(\ref{CoorSTM}) and a general solution of the homogeneous equation
\begin{equation}\label{STMHom}
(\hat L -1)\chi(x)=0.
\end{equation}
Physically, Eq.~(\ref{STMHom}) describes the atom-dimer channel just below the dimer breakup threshold ($\Psi_0 = 0$). 
Therefore, at distances $x\gg 1$, the function $\chi(x)$ is a linear combination of $\exp(ix)/x$ and $\exp(-ix)/x$.

In order to understand the short-distance behavior of $\chi(x)$, note the following property of the operator $\hat L$: 
\begin{equation}\label{property}
\hat L x^\nu =\lambda(\nu) x^{\nu-1}
\end{equation}
for any complex exponent $\nu$ in the region $-3<{\rm Re}(\nu)<1$, which is, in fact, the region of convergence of the integral 
on the left-hand side of Eq.~(\ref{property}) \cite{remark}. The function $\lambda(\nu)$ is given by
\begin{equation}\label{lambda}
\lambda(\nu)=-(\nu + 1)\tan\frac{\pi\nu}{2}-\frac{2\sin[\phi(\nu+1)]}{\sin(2\phi)\cos(\pi\nu/2)}
\end{equation}
and has two complex conjugate roots, $\nu_{1,2}=-1\pm is_0$, where $s_0$ is a real number satisfying Eq.~(\ref{s0equation}). 
At short distances the operator $\hat L$ in Eq.~(\ref{STMHom}) dominates over 1, and any solution of this equation should 
be a linear superposition of $\chi\propto x^{-1+is_0}$ and its complex conjugate. From now on we use the notation $\chi$ 
for the solution of Eq.~(\ref{STMHom}) with the following asymptotes:
\renewcommand{\arraystretch}{1} 
\begin{equation}\label{chi}
\chi(x)=\left\{\begin{array}{lr}Ax^{\nu}=Ax^{-1+is_0},&x\ll1\\
x^{-1}e^{ix+i\sigma-h}+x^{-1}e^{-ix-i\sigma+h},&x\gg1,
\end{array}\right.
\end{equation}
\renewcommand{\arraystretch}{1}where $A$ is a complex number, and $\sigma$ and $h$ are real numbers. The physical solution of 
Eq.~(\ref{STMHom}) (i.e., the one corresponding to a given three-body parameter) is expressed as
\begin{equation}\label{chi_physical}
\chi_{\theta}(x)=e^{i\theta}\chi(x)+e^{-i\theta}\chi^*(x),
\end{equation}
where $\theta$ is the three-body parameter (a complex number with imaginary part $\eta_*$).

The normalization in Eq.~(\ref{chi}) is chosen such that
\begin{equation}\label{orthonorm}
\langle p\chi(px)|p'\chi(p'x)\rangle\!=\!\!\int_0^\infty\!\!\! p\chi(px) p'\chi(p'x) x^2 dx=2\pi\delta (p-p').
\end{equation} 
The first equality in Eq.~(\ref{orthonorm}) is our definition of the scalar product (note the absence of the complex conjugation), 
and the second equality follows from the fact that $p\chi(px)$ and $p'\chi(p'x)$ are eigenfunctions of the symmetric operator 
$\hat L$ corresponding to the eigenvalues $p$ and $p'$. They are orthogonal for $p\neq p'$ and their scalar product in the 
vicinity of $p=p'$ can be worked out in the same way as in
Ref.~\cite{LLQ} (see Sec.~21 there). A simple change of the 
integration variable in Eq.~(\ref{orthonorm}) leads to the completeness condition
\begin{equation}\label{completeness}
\int_0^\infty p^2\chi(px)\chi(px')dp=2\pi\delta (x-x')/x^2.
\end{equation}

Equations~(\ref{orthonorm}) and (\ref{completeness}) allow us to construct the integral operator $(\hat L-1)^{-1}$ needed 
to solve Eq.~(\ref{CoorSTM}). In order to avoid problems with divergence of the corresponding integrals let us introduce 
an auxiliary function $g_0(x)$ related to $f(x)$ by
\begin{equation}\label{ftog_0}
f(x)=4\pi\Psi_0 [-1-\lambda(0)/x+\lambda(0)\lambda(-1)g_0(x)].
\end{equation}
Substituting this expression into Eq.~(\ref{CoorSTM}) and using Eq.~(\ref{property}), we find that $g_0(x)$ satisfies 
the equation $(\hat L-1)g_0(x)=x^{-2}$. Applying the operator $(\hat L-1)^{-1}$ to $x^{-2}$ we obtain the following particular solution:
\begin{eqnarray}\label{partsolution}
g_0(x)&=&\frac{1}{2\pi x}\int_0^\infty \chi(z)dz \nonumber\\
&\times&\left[\int_0^\infty \frac{\chi(y)y\, dy}{y-x-i0}-\frac{2\pi i x\chi(x)}{1-\exp(-2\pi s_0)}\right ],
\end{eqnarray}
where the first integral is defined as
\begin{equation}\label{sense}
\int_0^\infty \chi(z) dz=\lim_{\epsilon\rightarrow +0}\int_0^\infty \chi(z)z^\epsilon dz.
\end{equation}
The rule of going around the pole in the second integral and the numerical coefficient in front of the second term in the 
square brackets on the right-hand side of Eq.~(\ref{partsolution}) regulate the entry of $\chi(x)$, which can be arbitrary, 
into the particular solution $g_0(x)$. Using this freedom, we choose these parameters in such a way that $g_0(x)$ does not 
contain oscillating terms proportional to $x^{-1+is_0}$ at small $x$. Direct calculation shows that in the limit 
$x\rightarrow 0$ the right-hand side of Eq.~(\ref{partsolution}) equals $g_0(x)\approx [\int_0^\infty\chi(z)dz]^2/2\pi x$ 
to the leading order in $x$. On the other hand, according to Eq.~(\ref{property}), the same quantity in the same limit can 
be written as $g_0(x)=(\hat L -1)^{-1}x^{-2}\approx \hat L^{-1}x^{-2}=1/[\lambda(-1)x]$, which leads to the result
\begin{equation}\label{intchi}
\int_0^\infty\chi(z)dz=\sqrt{2\pi/\lambda(-1)}.
\end{equation}
Another consequence of our choice of the particular solution (\ref{partsolution}) is that removing the oscillating terms 
from $g_0(x)$ makes it real, since any imaginary part of $g_0$ would necessarily be a solution of the homogeneous 
Eq.~(\ref{STMHom}). Therefore, $g_0$ would have oscillations at short $x$, the absence of which we have ensured. 
Clearly, the function $f$ obtained by virtue of Eq.~(\ref{ftog_0}) is also real. Moreover, property (\ref{property}) 
ensures that $f=o(1)$ at small $x$ (i.e., its Taylor expansion starts with $x^1$ at least). Therefore, this solution of 
Eq.~(\ref{CoorSTM}) is not sensitive to the short-range physics and does not depend on the three-body parameter. 

Integrating Eq.~(\ref{partsolution}) in the limit $x\gg 1$ we get
\begin{equation}\label{g0largex}
g_0(x)\xrightarrow[x\rightarrow \infty]{}\frac{-i}{\sinh(\pi s_0)}\sqrt{\frac{2\pi}{\lambda(-1)}}\frac{\cos[x+\sigma+i(h+\pi s_0)]}{x}.
\end{equation}
It can be real only if $h=-\pi s_0$ [note that $\lambda(-1)<0$]; compare Ref.~\cite{Macek:2005}. 

Finally, the result that we are interested in is the linear combination
\begin{equation}\label{f_theta}
f_\theta (x)=f(x)+\gamma \chi_\theta (x),
\end{equation}
where the complex number $\gamma$ is chosen such that $f_\theta (x)$ contains only an outgoing wave at large $x$ 
(corresponding to an atom and a dimer flying apart after the three-body recombination event). This condition gives
\begin{equation}\label{beta}
\gamma=i\frac{\pi\Psi_0\lambda(0)\sqrt{2\pi\lambda(-1)}}{\sinh(\pi s_0)\cosh(\pi s_0-i\theta)}.
\end{equation}
Keeping only the relevant oscillating term at large $x$, we obtain
\begin{equation}\label{amplitude}
f_\theta (x)\xrightarrow[x\rightarrow \infty]{} i 4\gamma  \sin\theta \sinh(\pi s_0)\exp(ix+i\sigma)/x.
\end{equation}
Equation (\ref{amplitude}) together with Eqs.~(\ref{BethePeierls}) and (\ref{phi_b}) gives us the atom-dimer outgoing flux. 
Indeed, the large-$x$ asymptote $f=\xi \exp(ix)/x$, where $\xi$ is any complex amplitude, is accompanied by the flux $|\xi|^2\Phi_\infty$, where
\begin{equation}\label{Phi_infty}
\Phi_\infty=2\times (8\pi)^{-1} \times (4\pi) \times 2 = 2.
\end{equation}
Here we have explicitly written out the following factors: the factor of 2 reflects the two symmetric possibilities of forming 
the dimer (corresponding to the interchange ${\bf R_1}\rightleftarrows {\bf R_2}$), the factor of $(8\pi)^{-1}$ arises from the 
relation in between $\Psi$ and $f$ [see the discussion preceding Eq.~(\ref{phi_b})], the factor of $4\pi$ is the solid angle 
in the outgoing atom-dimer channel, and the last factor of 2 is the
atom-dimer relative velocity in the $x,y$ coordinates. 
The three-body recombination rate constant $\alpha_s$ is obtained by taking the squared modulus of the prefactor in front of 
$\exp(ix+i\sigma)/x$ in Eq.~(\ref{amplitude}) and by multiplying it by $\Phi_\infty$, by the factor of $1/2$, reflecting the 
fact that the number of pairs in the gas of species 2 is $n_2^2/2$, and by the factor $\hbar a^4/2\mu$ in order to restore 
the original physical units. We should also mention that the nine-dimensional volume $V^3$ is taken to be a unit volume 
 in the original system of coordinates $\{{\bf r,R_1,R_2}\}$. In the new coordinates $\{{\bf x}, {\bf y}, {\bf R}_{\rm c.m.}\}$, 
where ${\bf R}_{\rm c.m.}$ is the center-of-mass coordinate, this volume is $V^3=\cos^{-3}\phi$. The final result for the 
three-body recombination rate constant reads
\begin{eqnarray}\label{eq:alexact}
\hspace{-0.5cm}\alpha_s&=&\frac{32\pi^3\cos^3\!\phi\,\lambda^2(0)|\lambda(-1)|}{\sin\phi}\nonumber\\
&\times&\frac{\sin^2[s_0\ln(a/a_{*0})]+\sinh^2\eta_*}
{\sinh^2(\pi s_0+\eta_*)+\cos^2[s_0\ln(a/a_{0*})]}\frac{\hbar a^4}{m_1},
\end{eqnarray}
where we have expressed the three-body parameter $\theta$ through the original physical units:
\begin{equation}\label{theta}
\theta=s_0\ln(a/a_{*0})+i\eta_*.
\end{equation}
Formula~(\ref{eq:alexact}) is in excellent agreement with our numerical results.
Comparison with Eq.~(\ref{ashallow}) leads to \cite{remark2}
\beq\label{Cdelta}
C(\delta)=\frac{{(1+\delta)^2} \arcsin\left[1/(1+\delta)\right]-
\sqrt{\delta(2+\delta)}}
{2(4\pi-3\sqrt{3})}\,.
\eeq

As explained in Sec.~\ref{sec:results} the constant of the three-body recombination into deep dimers can be 
derived from the optical theorem. The result for $\alpha_d$ is given by Eq.~(\ref{adeep2}). Here we would like to show how 
one can derive this result by using the method of this section. In contrast to the recombination into shallow states we now 
have to look at the balance of the incoming and outgoing fluxes of atoms corresponding to the short-distance asymptote of 
$f_\theta (x)$ given by Eq.~(\ref{f_theta})
\begin{equation}\label{amplitude_0}
f_\theta (x) \xrightarrow[x\rightarrow 0]{} \gamma \chi_\theta (x)=\gamma (A e^{i\theta}x^{-1+is_0}+A^* e^{-i\theta}x^{-1-is_0}),
\end{equation} 
In analogy with $\Phi_\infty$, let $\Phi_0$ denote the number of atom triples disappearing at the origin (${\bf x}=0$, ${\bf y}=0$), 
provided the function $f$ takes the form of the incoming wave $x^{-1-is_0}$ with unit weight. With this definition, the 
recombination rate constant follows from Eq.~(\ref{amplitude_0}):
\begin{equation}\label{alpha_d}
\alpha_d=(\hbar a^4/2\mu) |\gamma|^2 \Phi_0 |A|^2 \sinh(2\eta_*).
\end{equation}
The product $\Phi_0 |A|^2$ can easily be found from definition (\ref{chi}) by equating the fluxes at $x\rightarrow 0$ 
and at $x\rightarrow \infty$ and using Eq.~(\ref{Phi_infty}):
\begin{equation}\label{fluxes}
\Phi_0 |A|^2=2 \Phi_\infty \sinh(2\pi s_0) =4\sinh(2\pi s_0).
\end{equation}
Substituting Eqs.~(\ref{fluxes}) and (\ref{beta}) into Eq.~(\ref{alpha_d}), we obtain Eq.~(\ref{adeep2}) exactly. We point out 
that it is in principle possible to obtain $\Phi_0$, and, therefore, $|A|^2$, by substituting the expression $f=x^{-1-is_0}$ into 
Eq.~(\ref{PsiThroughf}) and calculating the incoming flux from the
resulting wave function $\Psi$ in the six-dimensional 
configurational $\{{\bf x,y}\}$-space. 

Our analytical approach can also be used to derive $\alpha_d$ on the negative side of the resonance ($a<0$). In this case 
we use the units $\hbar=|a|=2\mu=1$, and some equations described above should be modified accordingly. In particular, 
Eq.~(\ref{CoorSTM}) reads
\begin{equation}\label{CoorSTMnega}
(\hat L +1)\tilde f(x)=4\pi \Psi_0
\end{equation} 
and we now introduce an auxiliary function $\tilde g_0$ related to $\tilde f$ by
\begin{equation}\label{ftog_0nega}
\tilde f(x)=4\pi\Psi_0 [1-\lambda(0)/x+\lambda(0)\lambda(-1)\tilde g_0(x)],
\end{equation}
where $\tilde g_0$ satisfies $(\hat L +1)\tilde g_0(x)=x^{-2}$. We write the solution in the form
\begin{equation}\label{partsolutionnega}
\tilde g_0(x)=\frac{1}{2\pi x}\int_0^\infty \chi(z)dz \int_0^\infty
\frac{\chi(y)y\, dy}{y+x}
\end{equation}
and integrating it in the small-$x$ limit we get the asymptote
\begin{equation}\label{fsmallxnega}
\tilde f(x)\xrightarrow[x\rightarrow 0]{} i \frac{2\pi \Psi_0 \lambda(0)\sqrt{2\pi \lambda(-1)}}{\sinh(\pi s_0)}Ax^{-1+is_0}.
\end{equation}
The function $\tilde f$ is a solution of Eq.~(\ref{CoorSTMnega}), but its oscillations at small $x$ do not have (in general) 
the correct phase imposed by Eq.~(\ref{chi}). This difficulty is resolved by observing that $\tilde f^*$ also satisfies 
Eq.~(\ref{CoorSTMnega}). The correctly behaving solution reads
\begin{equation}\label{f_thetanega}
\tilde f_\theta(x)=\frac{\exp(i\theta)\tilde f(x)+\exp(-i\theta)\tilde f^*(x)}{\exp(i\theta)+\exp(-i\theta)},
\end{equation}
and by subtracting the outgoing flux from the incoming one at small $x$ we obtain the result, cf. Eq.~(\ref{adeep}),
\begin{eqnarray}\label{alpha_dnega}
\hspace{-0.5cm}\alpha_d(a<0)&=&16\pi^3\cos^3\!\phi\,\lambda^2(0)|\lambda(-1)|\coth(\pi s_0)/\sin\phi\nonumber\\
&\times&\frac{\sinh(2\eta_*)}
{\cos^2[s_0\ln(|a|/a_{0*})]+\sinh^2\eta_*}\frac{\hbar a^4}{m_1},
\end{eqnarray}
Equation (\ref{alpha_dnega}) also gives the ratio 
\begin{equation}\label{analyticratio}
|a_-|/a_{0*}=\exp(\pi/2s_0).
\end{equation}
In other words, the maxima of $\alpha_s$ and $\alpha_d\, (a>0)$ and
the maxima of $\alpha_d\, (a<0)$ are symmetric with respect 
to the center of the Feshbach resonance.

Let us now return to the case $a>0$ and discuss some properties of atom-dimer collisions just below the dimer breakup threshold 
($E=-0$). Namely, by substituting the large-$x$ asymptote of $\chi$ into Eq.~(\ref{chi_physical}) one readily obtains the 
$s$-wave contribution to the scattering $S$-matrix 
\begin{equation}\label{Smatrix}
S_0=-e^{2i\sigma}\cosh(\pi s_0+i\theta)/\cosh(\pi s_0-i\theta)\,.
\end{equation}
From this expression, one can calculate the atom-dimer inelastic rate constant at zero total energy:
\begin{eqnarray}
&&\hspace{-1cm}\beta(E=-0)=\pi \cos^2\!\phi\cot\phi\nonumber\\
&&\hspace{-0.5cm}\times\frac{\sinh(2\pi s_0)\sinh(2\eta_*)}{\sinh^2(\pi s_0+\eta_*)+\cos^2[s_0\ln(a/a_{0*})]}
\frac{\hbar a}{m_1}\,,
\end{eqnarray}
see also Eq.~(\ref{eq:beta_e0}).
Remarkably, $\beta\, (E=-0)$ also reaches its maximum at $a=|a_-|$, and for small $s_0$ the peak 
can be quite narrow. We conjecture that this 
behavior is due to the Efimov state that crosses the atom-dimer threshold at $a=a_*$ and then exists as a scattering resonance.
Our numerical results support this conjecture.
We find that for mass ratios $\delta<3.475$ the resonance peak moves to values $a>a_*$, while it moves to 
values $a< a_*$ for $\delta> 3.475$. 
The resonance peak then hits the three-atom threshold at $a=|a_-|\exp(-\pi/s_0)$.
In Fig.~\ref{fig:beta2}, we have shown  $\beta$ for the specific case $\eta_*=0.1$ and $\delta=0.471$ corresponding to the K-Rb-Rb 
system observed in the Florence experiment~\cite{Barontini:2009}. The trajectory of the scattering resonance
is given by the double-dot-dashed line in the figure.
For smaller $s_0$ leading to a larger scaling factor, we conclude that the center of this Efimov scattering resonance 
can travel quite far from the value $a=a_*$ at $E=-B_d$ to the value $a=|a_-|\exp(-\pi/s_0)$ at $E=0$. 
We have verified this behavior numerically. It is thus worth 
investigating this scattering resonance in the vicinity of the atom-dimer threshold experimentally.     

%Finally, we would like to mention that the analytic method can be generalized to the case when species 2 is fermionic 
%in the Efimov regime, i.e. $\delta<0.0735$. One has to remember though that in this case the main angular momentum channel 
%is $l=1$ because of the fermionic (anti)symmetry. 

%%%%%%%%%%%%%%%%%%%%%%%%%%%%%%%%%%%%%%%%%
%Comparison to Experiment
%%%%%%%%%%%%%%%%%%%%%%%%%%%%%%%%%%%%%%%%%
\section{Comparison to experiment}
\label{sec:comparison}

There are several experiments on heteronuclear Bose-Bose and Bose-Fermi mixtures, to which our results are directly applicable 
(in the universal limit). In Table~\ref{tab:nr1}, we present the universal predictions for some combinations of alkali isotopes 
being investigated at the moment and interesting from the viewpoint of Efimov few-body physics. We sort them by the value of the 
scaling factor. 
%-----------------------------------------------------------------------------------------------------
\begin{center}
\begin{table*}[t]
\hfill{}
\begin{tabular}{|l||l|l|l|l|l|l|}
\hline
 &$^7$Li-Cs-Cs&$^6$Li-Rb-Rb&$^7$Li-Rb-Rb&$^{40}$K-Rb-Rb&$^{41}$K-Rb-Rb&Rb-$^{41}$K-$^{41}$K\\
\hline\hline
$\delta$	&0.053	&0.069	&0.081	&0.460	&0.471&2.21\\
$s_0$&1.850& 1.635&1.523&0.6536&0.6444&0.2462\\
$\exp(\pi/s_0)$ &5.465& 6.835&7.864&122.7 &131.0&348000\\
$a_*/|a_-|$&0.13 &0.16  &0.18 &0.51&0.52&0.91\\
%%$C(\delta)$& 1.09&0.853&0.0792 \\   %%old definition of C0
$C(\delta)$&0.072 &0.068&0.066&0.037&0.037&0.015 \\
$C_1(\delta)$&2.54 &2.33 &2.22&1.14& 1.13&0.94\\
$C_2(\delta)$&2.52 &2.5 &2.47& 2.08&2.07&1.30\\
\hline
\end{tabular}
\hfill{}
\caption{Universal parameters for various heteronuclear mixtures. The isotopes of rubidium and cesium are $^{87}$Rb and $^{133}$Cs.}
\label{tab:nr1}
\end{table*}
\end{center}
%-------------------------------------------------------------------------------------------------------

\subsection{The $^{40}$K-$^{87}$Rb mixture}
Zirbel {\it et al.}\ at JILA recently studied weakly bound fermionic $^{40}$K-$^{87}$Rb molecules and their stability in 
collisions with atoms near a wide (open-channel-dominated) heteronuclear Feshbach resonance at $B_0=546.7$~G \cite{Zirbel}. 
In particular, they measured the atom-dimer relaxation rate for collisions of these dimers with Rb atoms as a function of $a$. 
The corresponding data (see Fig.~2 in Ref.~\cite{Zirbel}) can be fit very well with our Eq.~(\ref{eq:beta}), where the fitting 
parameters are $a_*=200\pm 50\, a_{Bohr}$ and $\eta_*=0.05 \pm 0.02$. In the same work, the authors also measured the three-body 
recombination rate constant on both sides of this Feshbach resonance (i.e., in the same universal region). We fit their results 
with Eq.~(\ref{adeep}) on the negative side of the resonance and with the sum $\alpha_s+\alpha_d$ given by Eqs.~(\ref{ashallow}) 
and (\ref{adeep2}) on the positive side \cite{remark3}. A good agreement is achieved if we choose 
$a_*=300 \pm 100\, a_{Bohr}$ (for this mass ratio $a_-=-1.96 a_*$, see
Table~\ref{tab:nr1}) and the same $\eta_*$ as above. These parameter values lead to a peak of the three-body recombination at $a=a_- \approx -600\, a_{Bohr}$. Although in Ref.~\cite{Zirbel} 
the peak has not been identified, the overall shapes of $\beta$ and $\alpha$ measured for this particular Feshbach resonance 
indicate that it is worth performing a more detailed measurement of the three-body loss rate around this value of $a$.

\subsection{The $^{87}$Rb-$^{41}$K mixture}
The group of Inguscio and Minardi in Florence investigated a Bose-Bose mixture of $^{87}$Rb and $^{41}$K \cite{Barontini:2009}. 
They observed three loss resonances by scanning the scattering length and monitoring the population dynamics of the species in the 
vicinity of each of the resonances. For negative scattering length, they identified a K-Rb-Rb resonance at $a=-246\ a_{Bohr}$ and a 
K-K-Rb resonance at $a=-22000\ a_{Bohr}$. The third resonance is observed at the positive scattering length $a=667\ a_{Bohr}$ and 
attributed to enhanced atom-dimer scattering in the K-Rb-Rb three-body system. This process is assumed to contribute to three-body 
losses through multiple rescattering processes (see also Ref.~\cite{Zaccanti:2008}). An independent confirmation of this resonance in a 
system prepared directly out of K-Rb dimers and Rb atoms would be desirable. Unfortunately, in contrast to the JILA experiment, the 
dimers are bosonic and their short lifetime can make such a confirmation difficult~\cite{Minardi}. 

The interspecies van der Waals length in the K-Rb system is $\ell_{vdW}=72\, a_{Bohr}$, such that these resonances 
should be within 
the range of validity of the universal theory. Assuming that the observed K-Rb-Rb features are due to Efimov resonances, one can 
extract the ratio $a_*/|a_-|=2.7$ from the Florence experiment, whereas our theory predicts $a_*/|a_-|=0.52$. The discrepancy can 
be attributed to the effective range corrections. In particular, one should be careful with the feature at $a=-246\ a_{Bohr}$, which is not too large compared to the van der Waals length. 
Besides, if we believe in the ``rescattering'' nature of the positive-$a$ resonance, one should take into account a finite-energy 
shift of the position of the atom-dimer scattering resonance. Indeed, even at zero temperature, dimers formed by three-body 
recombination collide with stationary Rb atoms at the finite collision energy $[m_{Rb}^2/(m_{K}+2 m_{Rb})^2]B_d\approx 0.16 B_d$. 
In Fig.~\ref{fig:beta2}, we have shown numerical results for $\beta$ for energies from the
scattering threshold, $E=-B_d$, up to the breakup threshold, $E=0$, using $\eta_*=0.1$ and $\delta=0.471$ 
corresponding to the K-Rb-Rb resonance at $a=667\ a_{Bohr}$. At $E=0$, the resonance peak is 
only 7\% higher than the minimum value of $\beta$ and the resonance is almost completely washed
out. Moreover, the peak value of $\beta$ at $E=0$ is a factor of 300 smaller than at $E=-B_d$.
Of course, the explicit numbers depend on the value of  $\eta_*$, but
one should not exclude the possibility of other explanations of the positive-$a$ feature. 

%%%%%%%%%%%%%%%%%%%%%%%%%%%%%%%%%%%%%%%%%%%%%%%%%
\begin{figure}[t]
   %     \vspace*{0.4cm}
	\centerline{\includegraphics*[width=0.95\hsize]{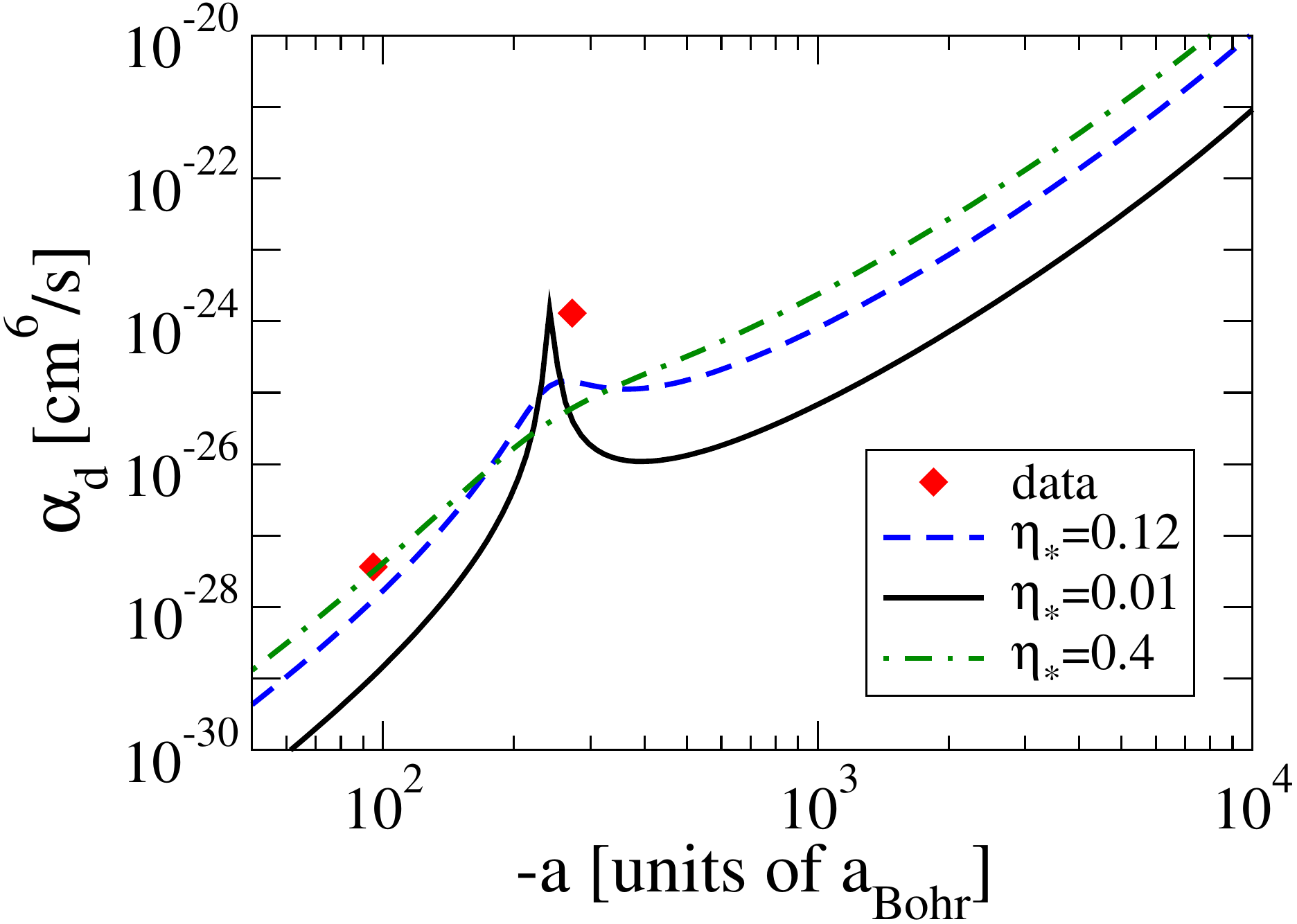}}
	\caption{(Color online) $\alpha_{d}$ vs $a$ for the $^{41}$K-$^{87}$Rb-$^{87}$Rb system assuming $a_-=-246\ a_{Bohr}$ and three values 
of $\eta_*$: $\eta_*=0.12$ (dashed line), $\eta_*=0.01$ (solid line), and $\eta_*=0.4$ (dash-dotted line). 
Data points indicated by diamonds are taken from Refs.~\cite{Barontini:2009, Minardi}; see text.}
	\label{fig:adeep}
\end{figure}
%%%%%%%%%%%%%%%%%%%%%%%%%%%%%%%%%%%%%%%%%%%%%%%%%
Aside from detecting the positions of the resonances, it is desirable to compare the actual shapes of the $a$-dependence of the 
loss rate with the theoretical predictions, especially on the positive side of the resonance, where $\alpha_s+\alpha_d$ is rather 
smooth. So far, the three-body recombination rate in the $^{87}$Rb-$^{41}$K mixture has been measured for two values of 
$a<0$~\cite{Barontini:2009,Barontini:err}, and the comparison with our calculation is rather inconclusive. Figure~\ref{fig:adeep} 
shows $\alpha_{d}$ for the K-Rb-Rb resonance at $a=-246\ a_{Bohr}$ where only the recombination into deep dimers can occur. The
dashed curve is calculated using the value $\eta_*=0.12$ suggested in Ref.~\cite{Barontini:2009}. The data point close to the 
resonance is taken from Ref.~\cite{Barontini:2009,Barontini:err}, whereas the one farther away from the resonance gives an upper 
limit of the recombination rate~\cite{Minardi}. In order to illustrate the sensitivity of the result to $\eta_*$, we also show 
it for $\eta_*=0.01$ (solid line) and  $\eta_*=0.4$ (dash-dotted line). The discrepancy between the measured 
recombination rate at the resonance and our result for $\eta_*=0.12$ is about one order of magnitude. In order to understand 
its origin, more measurements around the resonance position are required. Such data would allow for a more precise determination 
of the width parameter $\eta_*$ and of the resonance shape predicted by the universal theory. 

\subsection{Future experiments} 
The T\"ubingen group of Zimmermann recently studied the $^7$Li-$^{87}$Rb and $^6$Li-$^{87}$Rb mixtures. They identified and 
quantified several interspecies Feshbach resonances in both of them~\cite{Marzok:08,Deh:07} and have reached quantum 
degeneracy~\cite{Silber:05}. These mixtures are characterized by quite
small mass ratios and, therefore, small scaling factors, which is 
favorable for observing the discrete scaling invariance. Another very good candidate for studying the Efimov effect with even 
smaller scaling factors is a mixture of $^{133}$Cs with either isotope of lithium (the $^7$Li-$^{133}$Cs mixture was created 
in Heidelberg \cite{Deiglmayr:08}). The universal parameters for these mixtures can be found in Table~\ref{tab:nr1}.
Predictions for the three-body recombination and atom-dimer relaxation rates can be obtained
from Eqs.~(\ref{ashallow}), (\ref{adeep2}), (\ref{eq:beta}),
(\ref{eq:beta_e0}), and (\ref{adeep}).

%%%%%%%%%%%%%%%%%%%%%%%%%%%%%%%%%%%%%%%%%
%Conclusion
%%%%%%%%%%%%%%%%%%%%%%%%%%%%%%%%%%%%%%%%%
\section{Summary and conclusion}
\label{sec:sum}

In this paper, we have calculated the three-body loss rates in heteronuclear mixtures
of atoms for the case of large scattering length between the unlike atoms.
We have analyzed this problem using two complementary methods.

First, we have formulated a universal effective field theory  for this system and derived 
momentum-space integral 
equations for the trimer energies and the atom-dimer scattering amplitude. From an analysis of the 
bound-state equation we have calculated the ratio of the resonance positions $a_*/|a_-|$ as a function of the
mass ratio, $\delta$. Moreover, we have calculated the three-body recombination 
and atom-dimer relaxation rates numerically. We have  
provided semianalytical expressions for the rate constants of three-body recombination 
into shallow and deep dimers as a function of 
the interspecies scattering length $a$ and the Efimov width parameter $\eta_*$.
Furthermore, we have calculated the atom-dimer relaxation constant from the scattering threshold
at $E=-B_d$ up to the dimer breakup threshold at $E=0$.

Second, we have carried out an analysis of the recombination and relaxation process in configuration space.
We have generalized the method developed in Ref.~\cite{Petrov3Fermions} to heteronuclear 
bosonic mixtures and obtained analytic expressions for the recombination 
and relaxation rates at $E=0$.
We find excellent agreement of these expressions with our numerical results from the momentum-space
integral equations. 

The expressions in Eqs.~(\ref{ashallow}), (\ref{adeep2}),
(\ref{adeep}), and (\ref{Cdelta}) fully determine
the three-body recombination rates for heteronuclear bosonic mixtures with resonant scattering between the 
unlike atoms in the universal zero-range theory. The atom-dimer relaxation rates 
at $E=-B_d$ and $E=0$ are given by
Eqs.~(\ref{aADeq}), (\ref{eq:beta}), and (\ref{eq:beta_e0}). These equations are universal and can be 
used to analyze experimental data for 
any combination of atoms with the range applicability of the universal theory.

In Ref.~\cite{DIncao:05mass}, D'Incao and Esry give a general functional dependence of the recombination rates on the scattering 
length for all possible combinations of bosons and fermions. This includes the case of two identical bosons and a third atom 
with $J=0$ which
we address here.  We agree with their expressions for $\alpha_d$ in the case $a<0$ and for $\beta$. For $\alpha_s$, our general 
form (\ref{eq:alexact}) does not agree with their result. 
The proportionality of $\alpha_s$ to $\sin^2(s_0 \ln a +\phi_3)$ where $\phi_3$ is a short-range phase \cite{DIncao:05mass,DIncao:05} 
emerges only  if $\exp(2\pi s_0)\gg 1$ and the expression (\ref{eq:alexact}) can be simplified (cf.~Ref.~\cite{Braaten:2004rn}). 
This is the case for small mass ratios $\delta$.
Moreover, our prediction for the dependence of $\alpha_s$ on $\delta$ 
[see Eqs.~(\ref{ashallow}) and (\ref{Cdelta})] 
differs from the result $\alpha_s \propto [\delta(2+\delta)]^{3/2} a^4/ (1+\delta)^2 / m_1$ obtained in Ref.~\cite{DIncao:05mass}.

We have applied our results to some heteronuclear mixtures in ongoing and planned experiments. 
We find good agreement between theory and the JILA experiment~\cite{Zirbel}
that investigated $^{40}$K-$^{87}$Rb molecules and their stability in 
collisions with atoms near a wide heteronuclear Feshbach resonance at $B_0=546.7$~G.
For the recent experiment by the Florence group which uses
a mixture of $^{41}$K and $^{87}$Rb atoms \cite{Barontini:2009,Barontini:err}, we observe moderate
discrepancies between theory and experiment. We obtain $a_*/|a_-|=0.52$
for the resonance positions while  the experimental ratio is $a_*/|a_-|=2.7$. 
Because neither the effective range corrections nor the experimental errors
of the ratio are known accurately, no definite conclusion can be drawn at the moment. 
In particular, our analysis of atom-dimer relaxation suggests
that explanations should be considered other than an Efimov resonance
for the feature at $a=667\ a_{Bohr}$ that was used to extract the value of $a_*$.

Using the value $\eta_*=0.12$ extracted in Ref.~\cite{Barontini:2009}, 
we find that the calculated recombination rate
at the resonance is about one order of magnitude too small. 
Using smaller values of  $\eta_*$, the size of the 
experimental rate can be reproduced. 
In order to resolve this discrepancy, more measurements around the resonance position are required. Currently,
there are only two data points and $\eta_*$ cannot be determined accurately.
Additional data would allow for a more precise determination of $\eta_*$ and allow
for a test of the resonance shape predicted by the universal theory.

Finally, we have calculated the universal parameters determining the three-body 
loss rates for various other mixtures and have 
summarized them in Table \ref{tab:nr1}.
Extending earlier work by D'Incao and Esry \cite{DIncao:05mass,DIncao:05},
our predictions lay the theoretical ground for the experimental observation of Efimov physics 
in heteronuclear
mixtures. They should be useful for planning and analyzing future experiments.

\begin{acknowledgments}
We thank Eric Braaten, David Canham, and Francesco Minardi for discussions.
K.H. was supported by the \lq\lq Studien\-stiftung des
Deutschen Volkes'' and by the
Bonn-Cologne Graduate School of Physics and Astronomy.
H.W.H. acknowledges support from the the Bundesministerium f\"ur Bildung und Forschung, BMBF under Contract No.~06BN9006. 
D.S.P. is supported by the the $\hat{{\textnormal I}}$le-de-France
Cold Atom Research Institute, IFRAF, by the French National Research
Agency, ANR (Grant No.\ 08-BLAN-65), by the EuroQUAM-FerMix program, 
and by the Russian Foundation for Fundamental Research.
\end{acknowledgments}

%%%%%%%%%%%%%%%%%%%%%%%%%%%%%%%%%%%%%%%%%%%%%%%%
%Bibliography
%%%%%%%%%%%%%%%%%%%%%%%%%%%%%%%%%%%%%%%%%%%%%%%%


\begin{thebibliography}{99}
\bibitem{remark} In momentum space we have to work with an integral operator, the kernel of which 
is given by Eq.~(\ref{Kernel}) with $E=0$. This operator possesses a property similar to Eq.~(\ref{property}), but the region of convergence in that case is narrower.
\bibitem{remark3} Note that our rate constant $\alpha$ is half the three-body loss coefficient 
$K_3$ defined in Ref.~\cite{Zirbel}.
\end{thebibliography}
\end{document}